\documentclass[12pt]{article}

\usepackage{color}
\usepackage{graphicx}

\newcommand{\dd}{\mathrm{d}}
\newcommand{\rk}{\mathrm{k}}

\newcommand{\Max}{\mathrm{Max}\,}
\newcommand{\Av}{\mathrm{Av}\,}

\newcommand{\cT}{\mathcal{T}}

\usepackage{amssymb}
\usepackage{amsmath}

\newcommand{\rf}[1]{(\ref{#1})}
\newcommand{\beq}{\begin{equation}}
\newcommand{\eeq}{\end{equation}}
\newcommand{\bea}{\begin{eqnarray}}
\newcommand{\eea}{\end{eqnarray}}

\newcommand{\e}{\mbox{e}}

\newcommand{\Lam}{\Lambda}
\newcommand{\bt}{\beta}

\newcommand{\tht}{\theta}
\newcommand{\ep}{\varepsilon}

\newcommand{\kp}{\kappa}

\newcommand{\ra}{\rangle}
\newcommand{\la}{\langle}

\newcommand{\cD}{{\cal D}}

\newcommand{\cO}{{\cal O}}

\begin{document}

\begin{center}
\vspace{24pt}
{\large \bf Euclidian 4d quantum gravity with a non-trivial measure term}
\vspace{30pt}

{\sl J. Ambj\o rn}$\,^{a,b}$,
{\sl L. Glaser}$\,^a$,
{\sl A. G\"{o}rlich}$\,^{a,c}$ and 
{\sl J. Jurkiewicz}$\,^c$

\vspace{48pt}

$^a$~The Niels Bohr Institute, Copenhagen University\\
Blegdamsvej 17, DK-2100 Copenhagen \O , Denmark.\\
email: ambjorn@nbi.dk, glaser@nbi.dk, goerlich@nbi.dk\\

\vspace{10pt}

$^b$~Institute for Mathematics, Astrophysics and Particle Physics (IMAPP)\\ 
Radbaud University Nijmegen, Heyendaalseweg 135,
6525 AJ, \\
Nijmegen, The Netherlands 

\vspace{10pt}

$^c$~Institute of Physics, Jagiellonian University,\\
Reymonta 4, PL 30-059 Krakow, Poland.\\
email: jerzy.jurkiewicz@uj.edu.pl\\

\vspace{96pt}
\end{center}


\begin{center}
{\bf Abstract}
\end{center}

\noindent 
We explore an extended coupling  constant space of 4d regularized Euclidean
quantum gravity, defined via the formalism of dynamical triangulations.
We add a measure term which can also serve as a generalized higher 
curvature term and determine the phase diagram and the geometries 
dominating in the various regions. A first order 
phase transition line is observed, but no second order transition point 
is located. As a consequence we cannot attribute any continuum 
physics interpretation to the so-called crinkled phase of 
4d dynamical triangulations.

\newpage

\section{Introduction}\label{introduction}

The lattice regularization of geometries called Dynamical Triangulations 
(DT) provides us with a 
regularization of four-dimensional Euclidean quantum gravity within the realm 
of ordinary quantum field theory \cite{aj,am}. Presently we do not know 
if such a theory exists. Clearly, if the starting action
is just the Einstein-Hilbert action, the resulting theory 
has to be non-perturbatively defined since an expansion
of the Einstein-Hilbert action around a fixed background geometry 
leads to a non-renormalizable theory and since the continuum 
Euclidean Einstein-Hilbert action is unbounded from below. 
The asymptotic safety
scenario of Weinberg discussed general conditions which  
such a non-perturbative field theory should satisfy, using the 
Wilsonian renormalization group (RG) framework \cite{weinberg}. 
The central idea 
was that there should exist a non-Gaussian fixed point 
which would define the UV limit of the theory. Evidence for 
such a fixed point has been found both using the $2+\ep$ expansion \cite{kawai}
and the so-called exact or functional  
renormalization group equation (FRG) \cite{FRG}.

The so-called Regge version of the Einstein Hilbert action is a natural, 
geometric implementation of the action on triangulations.
Using this action in the DT approach one has two bare (dimensionless) 
lattice coupling constants related to the gravitational coupling 
constant $G$ and the cosmological coupling constant $\Lam$. In this 
coupling constant space one was looking for a phase transition point
which could be a  candidate for the proposed asymptotically safe fixed point.
A fixed point was found, but the corresponding phase transition
turned out to be of first order \cite{firstorder}. 
Usually, for critical systems on a 
lattice one can only associate continuum field theories to 
the fixed points if the transition is higher than first order.
This result was disappointing, but in a larger coupling 
constant space one would  expect to see transitions
where one could take a continuum limit. One can clearly add 
higher order curvature terms to the Einstein action in such 
a way  that the theory becomes renormalizable. It has been  
shown a long time ago that adding  $R^2$ terms to the action
would make the gravity theory renormalizable because the propagator would 
fall off like $1/k^4$ and thus improve the UV behavior of 
the theory \cite{stelle}. The problem with such a realization 
of renormalizability of  quantum gravity is that 
it is expected to correspond to a non-unitary theory when 
rotated back to Lorentzian signature, precisely because of 
the additional poles present in the propagator falling off like $1/k^4$.
However, in the context of the RG approach in the Euclidean 
sector, with infinitely many coupling constants, there should exist
a critical surface associated with such a theory. 
Refined perturbative treatments \cite{max} as well as 
the the use of FRG methods \cite{roberto,frank} provide evidence for this
by identifying a fixed point asymptotically free (i.e. Gaussian) 
in  coupling constants associated with the $R^2$ terms and asymptotically
safe in $\Lam$ and $G$. This fixed point seemingly differs from
the ``purely'' asymptotic safe fixed point discussed above, 
where also the coupling constants associated with the $R^2$-terms 
are different from zero \cite{frank}.

Since DT is a lattice regularization of Euclidean geometries it is 
natural to consider an enlarged  coupling constant space involving 
higher curvature terms. Such terms would most likely
be generated anyway if one could apply the Wilsonian RG 
techniques to the DT lattices. Similarly, being a lattice 
regularization, it has the potential to include 
the non-perturbative contributions alluded to above.
It has already been attempted to  explicitly include the higher curvature terms
in the DT formalism \cite{highcurve}.
The Regge action on a d-dimensional triangulation is defined as 
the sum of the deficit angles around the $(d-2)$-dimensional 
subsimplices times the $(d-2)$-dimensional ``volumes'' of these 
subsimplices. This gives a beautiful geometric 
interpretation to the Einstein action in $d$-dimensional spacetime
\cite{regge}. 
The DT formalism ``builds'' its $d$-dimensional triangulations 
from identical d-simplices where all links have the same length, $a$, the 
lattice spacing. For a given $(d-2)$-dimensional subsimplex $t_{d-2}$ let 
$o(t_{d-2})$ denote the {\it order} of $t_{d-2}$, i.e.\ the number of 
$d$-simplices to which $t_{d-2}$ belongs. The deficit angle of $t_{d-2}$ is  
\beq\label{1.1}
\ep(t_{d-2})  = 2\pi - o(t_{d-2}) \tht_d,~~~~\tht_d = \cos^{-1}(1/d).
\eeq
In two dimensions
we have $\tht_2=\pi/3$ and there is no intrinsic curvature when 
we glue together 6 equilateral triangles. Unfortunately 
there is no equally beautiful geometric realization of
higher curvature terms. The attempts to represent 
higher curvature terms naively as $\ep(t_{d-2})^2$ in 4d suffered from the 
problem that contrary to the situation in 2d, no flat spacetime can be build 
from gluing together the equilateral 4d building blocks used in DT.
While this does not exclude the possibility that this type of 
spacetimes could lead to sensible results when used in the path integral,
the end result of adding an $\ep(t_{d-2})^2$ term was as follows:
for a small coupling constant one found the same phases as without the 
$\ep(t_{d-2})^2$ term. For large coupling constants the system got stalled in 
weird configurations minimizing 
$\ep(t_{d-2})^2$, but having nothing to do with flat space. 
Somewhat more complicated and less local 
ways to implement $R^2$ terms are needed in the DT formalism,
but so far none that at the same time are useful for 
computer simulations have been found.  

However, evidence for a potentially non-trivial phase structure of 
DT came from another source, namely by changing the measure term
\cite{crinkled}.
The starting point of DT is the conjecture that the continuum 
path integral 
\beq\label{1.3}
Z = \int \cD [g]\,  \e^{- S^{EH}[g]}, 
\eeq
can be represented via a sum over simplicial manifolds built of 
equilateral four-simplices 
\beq\label{1.4}
Z =\sum_{\cT} \frac{1}{C(\cT)}\, \e^{- S^{R}[\cT]} .
\eeq
The symmetry factor $C(\cT)$ is the order of the 
automorphism group of a triangulation $\cT$.
The Regge version of the continuum Einstein-Hilbert action, 
\beq\label{1.4a}
S^{EH}[g] = - \frac{1}{G} \int \dd t \int \dd^D x \sqrt{g} (R - 2 \Lambda), 
\eeq
has a particularly simple realization in  DT  since
all four-simplices are identical and equilateral:
\beq\label{1.5}
S^R[\mathcal{T}] = - \kappa_2 N_2 +\kappa_4 N_4, 
\eeq
where $N_2$ is the number of triangles and $N_4$ the number of four-simplices.
The bare coupling constants $\kappa_2,\ \kappa_4$ are related to
the bare Newton's constant $G$ and the bare cosmological constant $\Lambda$,
respectively.

In the path integral \rf{1.4} each triangulation carries the 
same weight (except for the symmetry factor which is one for 
almost all triangulations). However even in the continuum 
it is somewhat unclear which measure $\cD [g]$ one should choose
for  the geometries. In the early history of  DT a number of 
different choices were suggested \cite{measure}, and 
in \cite{enzo} a 4d measure was proposed which 
contained  a  factor $\prod_{t=1}^{N_2} o_t^\beta$:
\beq\label{1.6}
\sum_{\cT} \frac{1}{C(\cT)} ~~\to~~ \sum_{\cT}\frac{1}{C(\cT)} \;
\prod_{t=1}^{N_2} o_t^\beta,
\eeq  
 where $o_t$ is the order of triangle $t$. In 2d Euclidean 
quantum gravity, regularized by DT, one can add a similar 
term, only replacing triangles in \rf{1.6} with vertices.
Both in 2d and 4d \rf{1.6} would then refer to 
$(d-2)$-dimensional subsimplices and via \rf{1.1} to higher 
curvature terms, although the identification is rather 
indirect and to a series of higher curvature terms. 
From a renormalization group point of view it 
should not be that important, since
one is just looking for a new fixed point with different physics. 
It was eventually  shown in \cite{kazakov} that the continuum limit 
of the 2d lattice theory was independent of any reasonable 
choice of $\bt$ in \rf{1.6}. The interpretation given in 
2d was that higher curvature terms were irrelevant operators
in a renormalization group framework (which is true from a naive 
power counting point of view). In 4d we do not 
have analytical results and it is possible that
the choice of weight factor {\it is} important for a continuum limit\footnote{
The interesting paper \cite{dario} presents a model 
which has an effective measure term similar to \rf{1.6} and 
where it actually {\it is} possible to perform some analytic 4d calculations. 
Unfortunately it is not  clear how closely related the model 
is to the DT models considered in this article. Nevertheless, in this 
model the measure term can change the phase structure.},
and that if this was the case, the 
choice \rf{1.6} could be viewed as some effective 
representation of higher curvature terms. The implementation
of the higher curvature terms via \rf{1.6} is less direct
then the native (and failed) attempt to simply add $\ep^2(t)$
from \rf{1.1}, as mentioned above.

In \cite{crinkled} it was observed that one seemingly 
entered a phase dominated by a new kind of 
geometries, named the ``crinkled phase'' by choosing $\bt$
sufficiently negative. The fractal dimension (the 
Hausdorff dimension) of typical geometries was reported
close to 4 and the spectral dimension around 
1.7. Potentially this new phase could reflect the 
presence of higher curvature terms 
and thus also, according to the FRG results \cite{frank}, a 
non-Gaussian asymptotically safe fixed point.
 
Interestingly,  the same phase was observed
when coupling gauge fields to gravity in four dimensions 
\cite{crinkled,ckr,yukawa}.
This was in contrast to the situation for a 
scalar field coupled to gravity, where little change 
was observed. However, the reported difference between scalar fields 
and gauge fields coupled to 4d gravity could be understood 
as a consequence of a different choice of discretized 
coupling of matter to the (piecewise linear) geometry. If the 
gauge fields were coupled in the same way as the scalar
fields the back reaction was equally weak as 
reported for scalar fields. The difference amounted to placing the 
gauge fields on the triangles of the 4d triangulation or placing 
them on the so-called dual triangles. It is possible to show that
a transformation between the two setups leads to a weight factor 
of the form \rf{1.6}. This gave some arguments in favor of 
viewing the crinkled phase as a lattice artifact, since one 
would not think it should make a significant difference if 
one used the lattice
or the dual lattice for the gauge fields \cite{aak}. However, it is fair to
say that the situation was unsettled, with some people claiming
that the crinkled phase represented continuum physics \cite{yukawa}.

Recently, there has been a renewed interest in the crinkled phase
after it was observed that the spectral dimension in the crinkled 
was scale dependent \cite{Laiho} and seemingly behaved more or less like 
the spectral dimension in so-called Causal Dynamical Triangulations 
(CDT)\cite{spectralcdt}. CDT is an attempt to formulate a theory of
quantum gravity where the path integral includes only geometries 
which allow a time foliation (see \cite{physrep} for a review). 
Such a foliation constraint can
best be motivated starting out in spacetimes with Lorentzian 
signatures, which is how CDT was originally formulated. However, for the 
purpose of numerical simulations the time direction has been rotated
such that the spacetimes studied on the computer have Euclidean 
signature. The result was a different phase structure compared 
to the one observed using DT, in particular it 
includes a second order phase transition line where 
one might be able to define a continuum limit. 
This is in principle a desirable situation, and the results in 
\cite{Laiho} for the spectral dimension open up the possibility 
that the crinkled phase could be identified with the so-called 
``phase C'', in the CDT phase diagram.  

A priori one can not rule out such an identification\footnote{There 
are also other possible interpretations of the continuum 
limit of the CDT theory, in particular that it can be related to 
Ho\v{r}ava-Lifshitz gravity \cite{hl}. For a detailed discussion we refer 
to the review \cite{physrep}.}. The geometries
which enter in the path integral in CDT after rotation to 
Euclidean signature are a subset of  those used in DT and 
effectively this restriction could move the theory into 
the same universality class as the theories with higher curvature 
terms, i.e.\ (again relying on the FRG picture) into the universality
class corresponding to the standard  asymptotic safety scenario.
This would have an interesting implication. One can show that
the CDT theory is unitary (it has a reflection positive transfer 
matrix related to the lattice time foliation \cite{cdttransfer}) and in this 
way it would add arguments in favor of the putative 
asymptotic safety theory actually 
being unitary, a fact which  is not obvious.   

In the following we investigate the effects of 
modifying the  measure term in the way displayed in eq.\ \rf{1.6}. 
   
\section{The numerical setup}\label{setup}

Viewing the modification of the measure term as part of the action, 
our action now depends on three bare coupling constants 
$\kp_2$, $\kp_4$ and $\bt$. In our simulations $\kp_4$ is not 
really a coupling constant since 
we  keep $N_4$, the number of four-simplices, (almost) fixed. 
More precisely we work in  
a pseudo-canonical ensemble of manifolds with topology $S^4$,
and use the partition function 
\beq\label{2.1}
 Z(\kappa_2, \kappa_4, \beta) = 
\sum_\mathcal{T} \frac{1}{C_\mathcal{T}} 
\cdot \prod_{t=1}^{N_2} o_t^\beta \cdot 
\e^{-\left[-\kappa_2 N_2 + \kappa_4 N_4 + \varepsilon (N_4 - 
\bar{N}_4)^2 \right]}.
\eeq
The quadratic term proportional 
to $\varepsilon$ fixes the total volume 
around some prescribed value $\bar{N}_4$.
To achieve this the bare cosmological constant has to be 
tuned to its critical value $\kappa_4 \approx \kappa_4^c$, the critical 
value being the value below which the partition function is
divergent. 

We use Monte Carlo simulations to study expectation 
values of observables in the ensemble defined by the 
partition function \rf{2.1}. The set of triangulations of 
$S^4$ we use are the so-called combinatorial triangulations,
where every $4$-simplex is uniquely defined 
by a set of $5$ distinct vertices and by demanding that  two 
adjacent $4$-simplices share precisely one face (a three-dimensional 
subsimplex). This is in contrast to the degenerate 
triangulations, defined in  \cite{Degenerate}, and 
used in the recent study of the crinkled phase \cite{Laiho}.  
It is believed that the models defined by combinatorial triangulations
and degenerate triangulations belong to the same universality class,
and using a different class of triangulations than used 
in \cite{Laiho} will give us a check of the robustness of the results 
obtained in \cite{Laiho} as well as in this study.

In the Monte Carlo simulations we use the standard 5 Pachner moves to 
update the four-dimensional combinatorial triangulations. 
For $d$-dimensional combinatorial triangulations of fixed 
Euler number the $d+1$ Pachner moves are local changes of the triangulations 
which are ergodic \cite{varsted}. 

Thus we will be exploring the coupling constant space ($\kp_2,\bt$).
We will use Monte Carlo simulations to generate a number of independent 
configurations for each value of $\kp_2$ and $\bt$ in a grid in the ($\kp_2,\bt$)-plane with $\bt$ between $0$ and $-2 $ varied in steps of $\delta \bt=0.2$ and $\kp_{2}$ between $0.5$ and $1.5$ varied in steps of $\delta \kp_{2}=0.1$. 
Using these we will calculate the expectation values 
of observables  $\cO$ over these configurations:
\beq\label{2.2}
\la \cO\ra_{conf} = \frac{1}{N_{conf}} 
\sum_{i=1}^{N_{conf}}  \cO_i,
\eeq
where $N_{conf}$ denotes the number of Monte Carlo generated 
independent configurations at a particular value of coupling 
constants and $\cO_i$ denotes the value of the observable $\cO$ 
calculated for the $i^{th}$ configuration, $i=1,\ldots,N_{conf}$.


\section{The phase diagram}\label{phases}

In order to determine the phase structure of the model
we measured a number of  ``observables'' which 
can be used to characterize  the geometries 
in the different phases. Observables which have in the past been 
useful in distinguishing between the two phases observed for $\bt=0$
include  the average number of vertices 
$\langle N_0 \rangle$ and the average number of 
triangles  $\langle N_2 \rangle$, as well as their associated 
susceptibilities 
\beq\label{3.0}\chi( N_0 ) \equiv 
 \frac{\langle N_0^2 \rangle - \langle N_0 \rangle^2}{ N_4},
~~~~ 
\chi( N_2 ) \equiv  \frac{\langle N_2^2 \rangle - \langle N_2 \rangle^2}{N_4}. 
\eeq

Another observable which will be useful is the radius volume profile $V(r)$.
We define and measure it as follows. Given two four-simplices 
we define a path between these as a  piecewise linear 
path between centers of neighboring four-simplices, connecting 
the centers of the two four-simplices. The (graph) {\it geodesic distance} 
between the two four-simplices is defined as the smallest number of 
steps in the set of paths connecting them.
For a given configuration and an initial simplex $i_0$,
the number of four-simplices  at a geodesic distance $r$ 
from $i_0$ is denoted as $V(r, i_0)$.
The average over configurations and initial points is then given by
\beq\label{3.1} 
V(r) \equiv \langle \frac{1}{N_4} \sum_{i_0} V(r, i_0) \rangle_{conf}. 
\eeq
The average radius is then defined as 
\beq\label{3.2}
\langle r \rangle \equiv \frac{1}{N_4} \sum_r r \cdot V(r). 
\eeq



We also look for the presence of so-called \emph{baby universes} 
separated by minimal necks.
A minimal neck is a set of five tetrahedra,
connected to each other, and forming a $4$-simplex
which is not present in the triangulation.
Cutting the triangulation open along the five 
tetrahedra will separate the triangulation in two 
disconnected parts, each with a boundary consisting 
of the five tetrahedra, the minimal boundary possible 
for the class of triangulations we consider. 
The analysis of baby universe distributions has been very 
useful as a tool to distinguish various phases of 
different geometries in 4d simplicial quantum gravity \cite{4dbaby},
as well as in the studies of 2d quantum gravity \cite{2dbaby}.

\subsection{Grid and phase diagram}

\begin{figure}
\begin{center}
\includegraphics[width=0.49\textwidth]{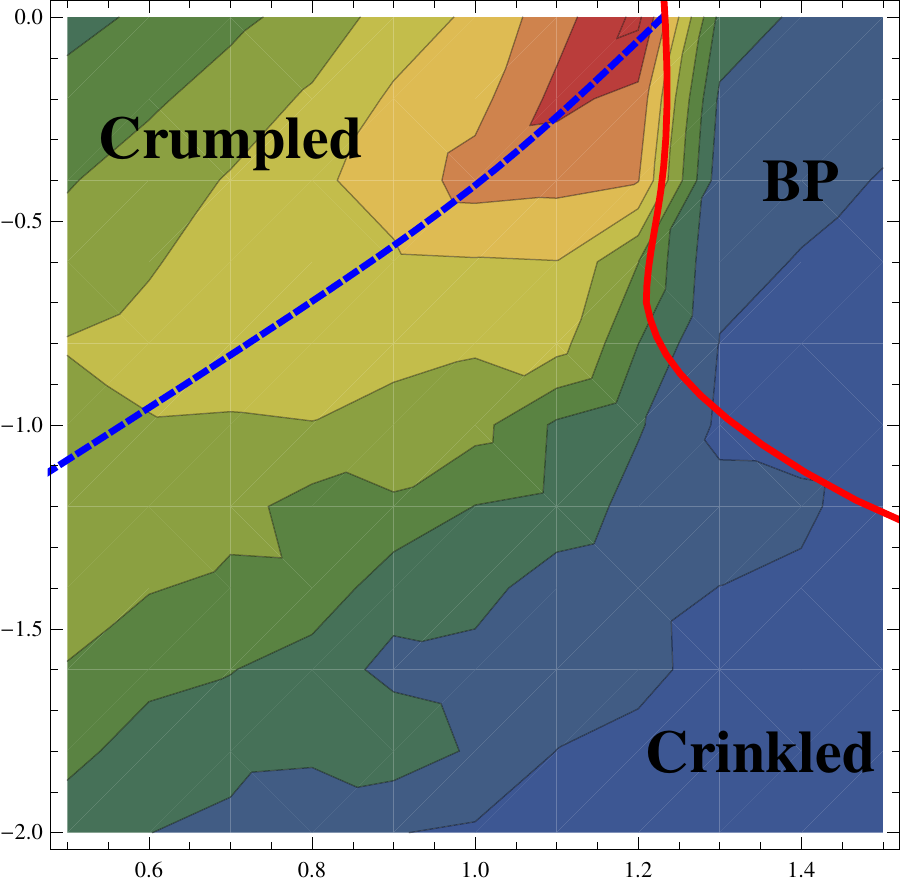}
\includegraphics[width=0.49\textwidth]{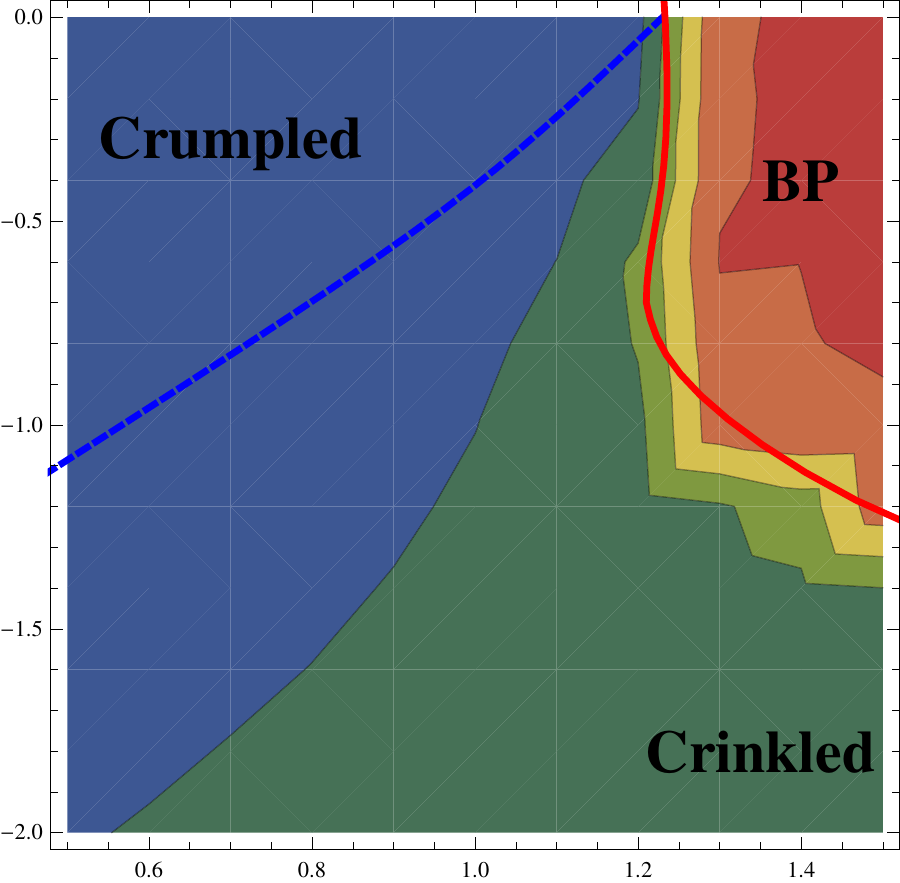}
\end{center}
\caption{Density plots of the susceptibility $\chi(N_0)$ (left) and the average radius (right) in the $(\kappa_2,\beta)$ plane for 
$\langle N_4 \rangle = 160 000$}
\label{fig:grid}
\end{figure}

\begin{figure}
\begin{center}
\includegraphics[width=0.32\textwidth]{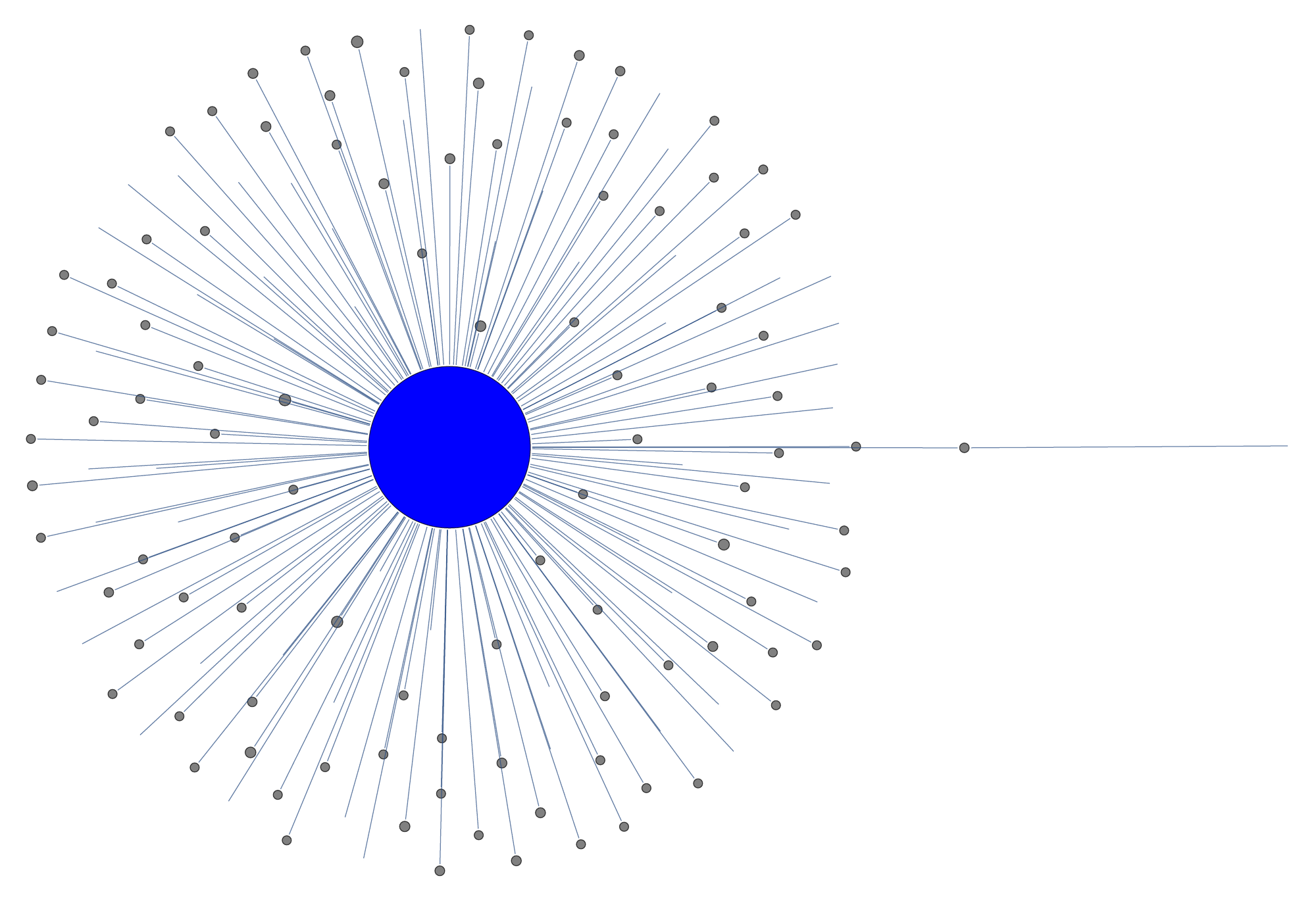}
\includegraphics[width=0.32\textwidth]{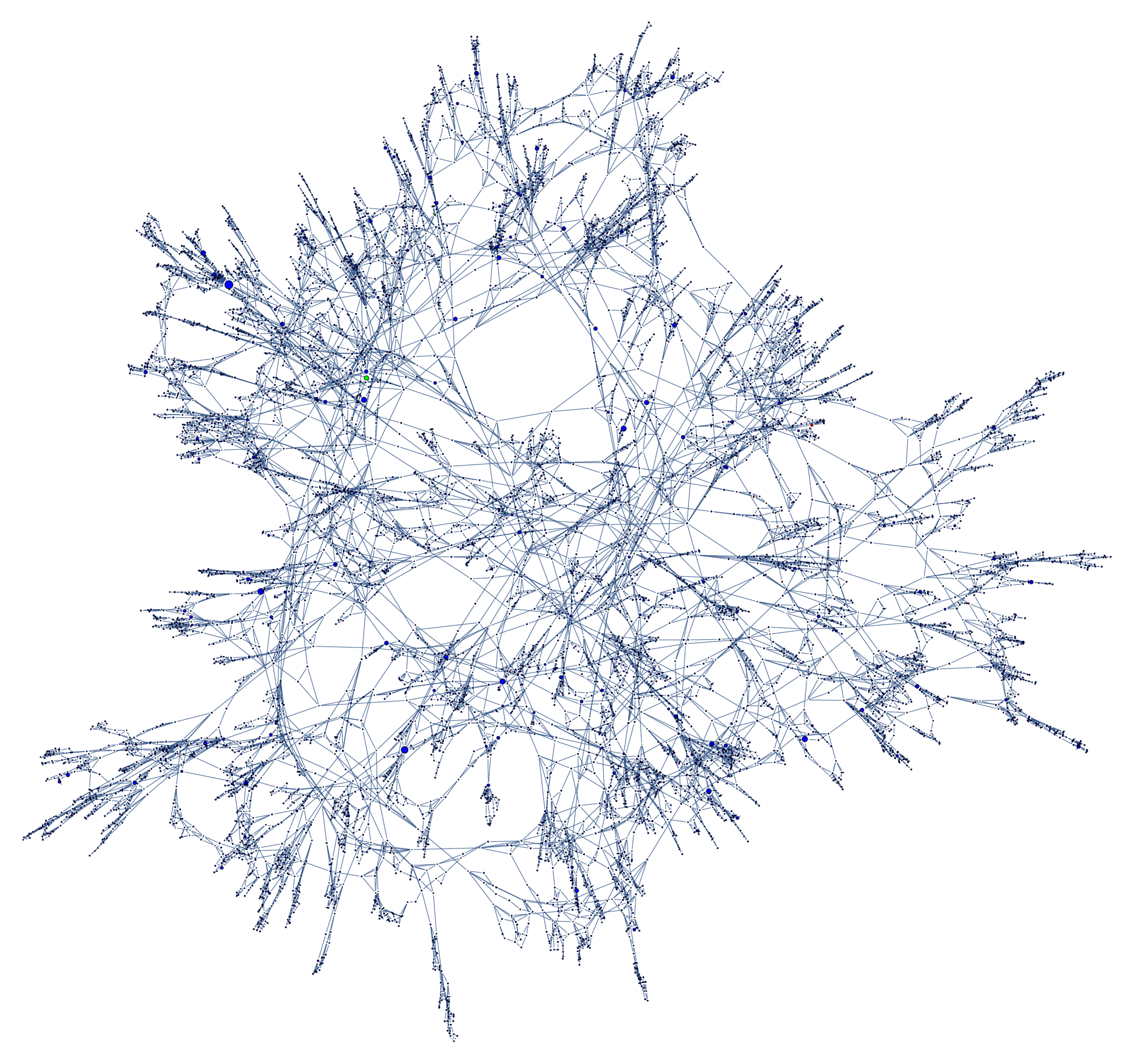}
\includegraphics[width=0.32\textwidth]{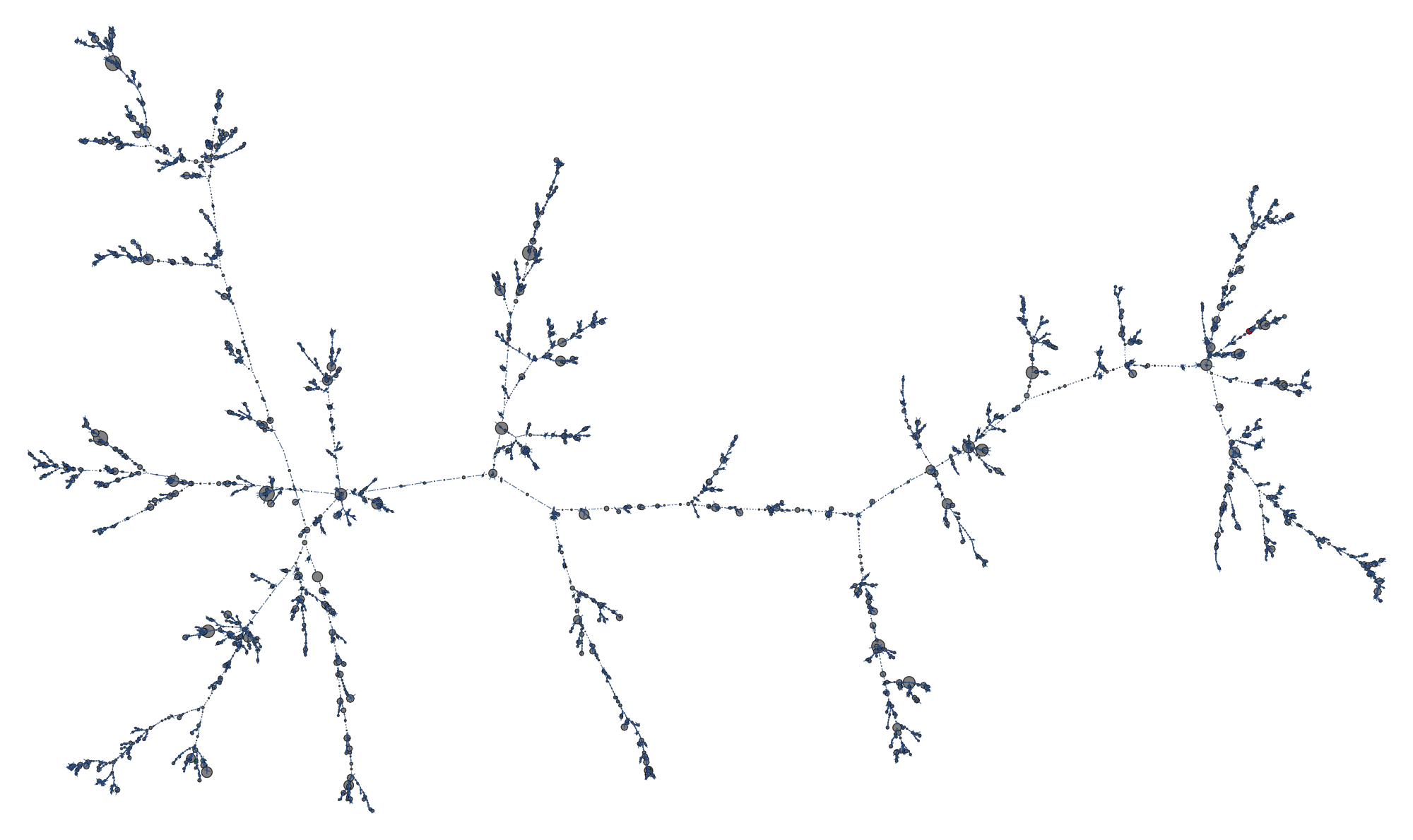}
\end{center}
\caption{Minimal baby universe graph of a typical configuration in, 
respectively from left to right, 
the crumpled phase, the crinkled region and the branched polymer phase.}
\label{fig:tree}
\end{figure}

In the case without non-trivial  measure term, i.e.\ when $\beta = 0$, 
there exist only two phases, namely the \emph{crumpled} phase 
and the \emph{branched polymers} phase \cite{aj,am,aj1,ckr1}. However, 
they are separated by a first order transition \cite{firstorder},
as already mentioned. It occurs at $\kappa_2 \approx 1.29$. 
At this point, we observe a 
peak in both  susceptibilities $\chi(N_0)$ and  $\chi(N_2)$,
as well as  a jump in $\langle r \rangle$. There is also an abrupt 
change in the baby universe structure 
as depicted in Fig.\ \ref{fig:tree}. The 
left graph in Fig.\ \ref{fig:tree} shows 
the baby universe structure for a typical 
configuration in the crumpled phase. 
One has a huge ``parent'' universe decorated 
with almost minimal baby universes (which are really too small to  
deserve being called (baby)-universes). The situation is quite the opposite
in the branched polymer phase, as shown on the right graph in 
Fig.\ \ref{fig:tree}. In this phase one has a genuine fractal structure 
of baby universes of all sizes. From 
a continuum point of view the problem with this phase is that the spacetime {\it is} too 
fractal, and spacetime itself, not only the baby universe structure,
seems to be described as a 2d fractal tree\footnote{The only 
exception might be very close to the transition point where arguments 
have been given in favor of a different interpretation of 
the fractal structure \cite{js}.}.

The additional coupling constant $\beta$ may introduce new phase(s).
We have extensively investigated a grid of points in the  
$(\kappa_2,\beta)$ plane, including  
the transition point $\beta = 0, \kappa_2 \approx 1.29$.
Plots of the susceptibility $\chi(N_0)$ (left) and the average radius (right) 
for the grid points are shown in Fig. \ref{fig:grid} 
($\kappa_2$ - horizontal axis, $\beta$ - vertical axis).
For negative $\beta$ the maximum of variance $\chi(N_0)$ (blue line) 
and a jump in $\langle r \rangle$ (red line) do not coincide any more.

It is observed that the branched polymer phase corresponds to large values of 
$\langle r \rangle$ and a {\it jump} to smaller values of the 
expectation value is very clear when 
one leaves the branched polymer phase. In this sense 
the branched polymer phase can be clearly distinguished 
from other phases by the red curve in Fig.\ \ref{fig:grid}.
The (not very pronounced) peak in the  susceptibility 
seems not to be a signal of a phase transition, as we will 
discuss later.

We also observe a region in coupling constant space where
the properties of typical configurations are in between those of the
crumpled phase and the branched polymer phase. It is natural to 
try to classify configurations in this region  as being in
the  hypothetical new crinkled phase.
This region is located around the point $\kappa_2 = 2.0, \beta = -2.0$ 
\footnote{For such values of the coupling constants 
the acceptance rate in the Monte Carlo simulations is relatively low,
and simulations take a painfully  long time.}.
The minimal baby universe structure is shown in Fig.\ \ref{fig:tree}.
Let us explain how the graphs shown there are constructed. We look for 
minimal necks. As already remarked a minimal neck consists of the 
five tetrahedra forming the boundary of a four-simplex, but such 
that the four-simplex is not part of the triangulation. We can cut
the triangulation in two disconnected parts along the five tetrahedra,
In this way we obtain two triangulations, each with a minimal boundary
(the five tetrahedra, now belonging to both triangulations). For each
triangulation we now repeat this process of finding baby universes and 
in this way we end up  with a number of disconnected universes with boundaries.
We represent each universe with a dot and we connect the dots by a link
if their boundaries had originally shared at least one  tetrahedron.
In this way minimal necks naturally equip triangulations 
with graph structures like the ones shown in Fig.\ \ref{fig:tree}.

In the crumpled and branched polymers phases
it happens very seldom that two minimal necks are neighbors. In these 
phases the graphs are thus tree graphs, bearing in mind that 
the topology of spacetime is that of $S^4$.

The situation is different in the crinkled region.
In this region we observe triangles of high order. 
We observe that a number of the tetrahedra  sharing such a triangle 
can belong to two minimal necks. In this way the graph
can contain a (long) loop ``twisted'' around a high order triangle. 
Such loops spoil the tree structure seen in the crumpled and branched polymer
phases.

For configurations belonging to the crumpled or 
the branched polymer phases we never observe triangles of high order,
while in the crinkled region the maximal order of triangles 
seems to behave like $\langle \Max o_t \rangle \propto N_4^{1/6}$.
At a first glance one would expect that  the measure term,
\beq\label{3.3}
\prod_{t=1}^{N_2} o_t^\beta = e^{\beta \cdot \sum_t \log o_t} 
\eeq
would suppress  high order triangles for negative $\beta$. What 
really happens is that the value of the observable  conjugate to 
$\bt$, i.e.\ $\langle \sum_t \log o_t \rangle$, indeed decreases
with decreasing $\beta$. However, the distribution of 
triangles-order $P(o_t)$ has a long tail and this makes it 
possible that even with a decreasing $\langle \sum_t \log o_t \rangle$ 
we can have an increasing 
$\langle \Av o_t \rangle$ and $\langle \Max o_t \rangle$, which is 
what we observe.

When we move from the branched polymer phase to the crinkled phase
the baby universe structure changes relatively smoothly. However,
as mentioned above, the transition between the two phases 
is seen clearly by a jump in $\langle r \rangle$. At the same 
time one also observes  a (small)  peak in $\chi(\log o_t)$ 
(see Fig.\ \ref{fig:pathlogot}).

We also measured points outside of the grid region - 
in a less systematic way - 
and the results agree with the picture presented above.

\vspace{2ex}
\noindent Below we summarize characteristics for  
typical configurations from the branched polymer phase, the crumpled 
phase and the hypothetical crinkled region.

\vspace{2ex}
{\noindent \bf The branched polymers phase:}
\begin{itemize}
\itemsep=0.5mm
\item Elongated geometry, $\langle r \rangle \propto N_4^{1 / 2}$
\item Dominated by minimal necks separating baby universes
\item Probability of baby universe of size $V$ is 
$P(V) \propto V^{\gamma-2} (N_4 - V)^{\gamma-2},$
where $\gamma = \frac{1}{2}$ is the string susceptibility exponent.
\item Hausdorff dimension $d_h = 2$, spectral dimension $d_s = 4 / 3$
\item Tree-like structure (cf. Fig. \ref{fig:tree})
\end{itemize}
\vspace{2ex}
{\noindent \bf The crumpled phase:}
\begin{itemize}
\itemsep=0.5mm
\item Collapsed geometry, $\langle r \rangle$ grows slower than any
$N_4^\alpha$, $\alpha >0$.
\item Two singular vertices of order $o_v \propto N_4$ connected by a singular  
link of order $o_l \propto N_4^{2/3}$.
\item No baby universes beyond the size of a few four-simplices.
Thus no susceptibility exponent $\gamma$ (formally $\gamma= -
\infty$). 

\item Hausdorff dimension $d_h= \infty$, spectral dimension $d_s$
infinite or at least large.

\end{itemize}
\vspace{2ex}
{\noindent \bf The crinkled region}
\begin{itemize}
\itemsep=0.5mm
\item The properties interpolate between crumpled and branched polymer regions
for finite volume, but seem in most cases to approach those of 
the crumpled phase with increasing volume. While  $\langle r \rangle$
is larger than in the crumpled phase it still grows very slowly with $N_4$.

\item One observes triangles of high order,  proportional 
to  $N_4^{1/6}$, contrary to the situation in the crumpled and 
branched polymer regions.

\item Many baby universes, but no {\it large} baby universes and 
thus no finite string susceptibility $\gamma$ (formally $\gamma =-\infty$).

\item The baby universes define a tree-like structure, but this 
structure contains loops related to the triangles of high order
(see Fig.\ \ref{fig:tree}).

\item The Hausdorff dimension $d_h$ is large (most likely infinite) and 
the spectral dimension $d_s$ seems also large (growing with volume as 
far as we can check)

\end{itemize}

\subsection{The path in the  $\boldsymbol{(\bt,\kp_2)}$
plane}

\begin{figure}[h]
\begin{center}
\includegraphics[width=0.8\textwidth]{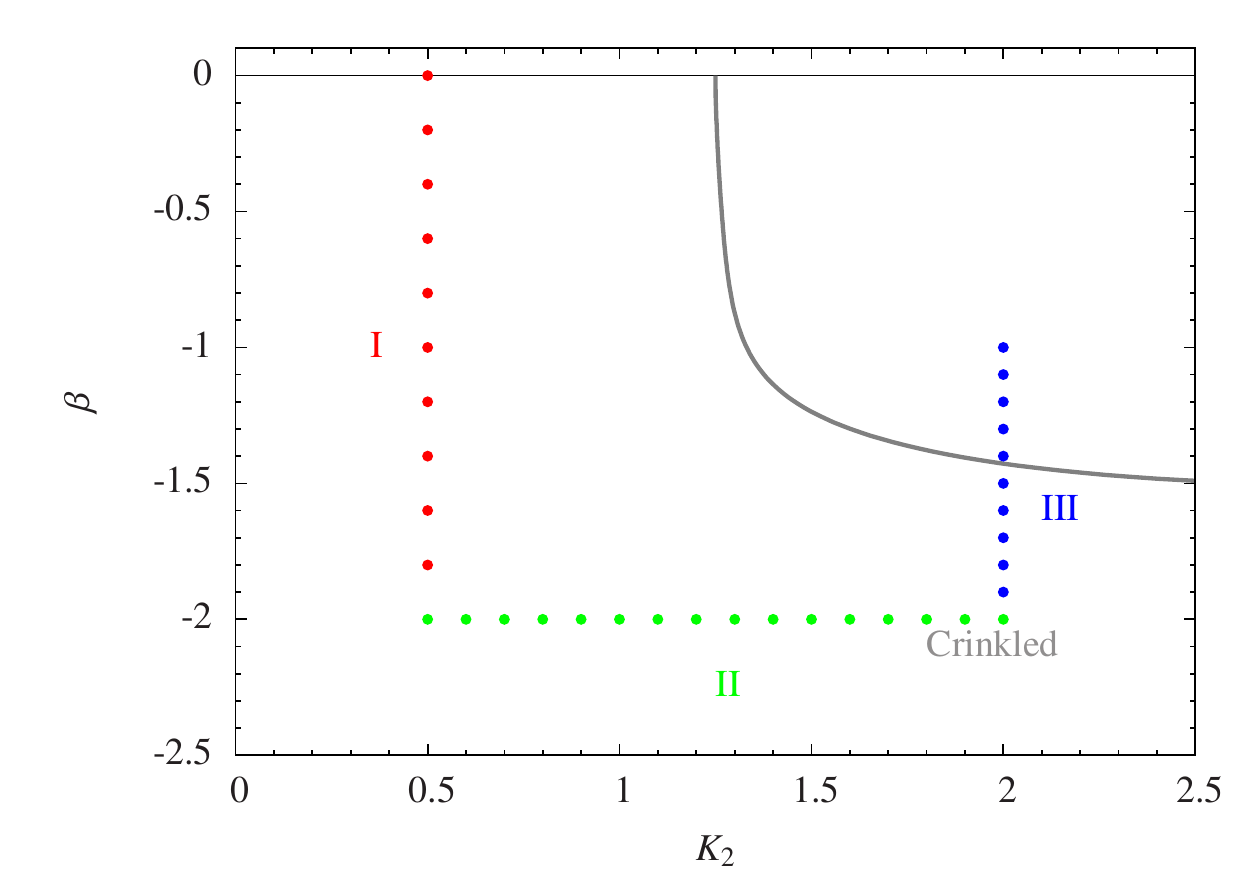}
\end{center}
\caption{A tentative phase diagram 
and a path (color points) from crumpled phase - through crinkled region - to branched polymer phase.
The thick gray line denotes the phase transition between branched polymers and other phases,
based on the grid measurements.}
\label{fig:phasediagram}
\end{figure}

In order to determine if there exists a  new  \emph{crinkled} phase
we need to perform simulations for various total volumes and 
check scaling of the observable.
Because this demands vast CPU resources, 
we follow the one-dimensional path shown in Fig.\ \ref{fig:phasediagram}
instead of using a full grid.
 
We performed measurements for three values of the 
total volume $N_4 = 40\rk$,  $80\rk$ and $160\rk$.
The path starts at a point in the crumpled phase 
$(\kappa_2 = 0.5, \beta = 0.0)$ and 
continuously leads through the 
crinkled region $(\kappa_2 = 2.0, \beta = -2.0)$
to stop in the branched polymers phase $(\kappa_2 = 2.0, \beta = -1.0)$.
If there is a phase transition between a crumpled and a crinkled phase,
the path will have to cross it.

The path consists of three segments
 marked with different colors to simplify comparison of plots:
a \textcolor{red}{red} vertical segment I at $\kappa_2 = 0.5$, 
a \textcolor{green}{green} horizontal segment II at  $\beta = -2.0,$ and 
a \textcolor{blue}{blue} vertical segment II at  $\kappa_2 = 2.0.$
We now describe the behavior of the various observables when 
we move along this path. 
  

\subsection{$\boldsymbol{N_0}$ and $\boldsymbol{N_2}$ observables}

\begin{figure}[h]
\begin{center}
\includegraphics[width=0.49\textwidth]{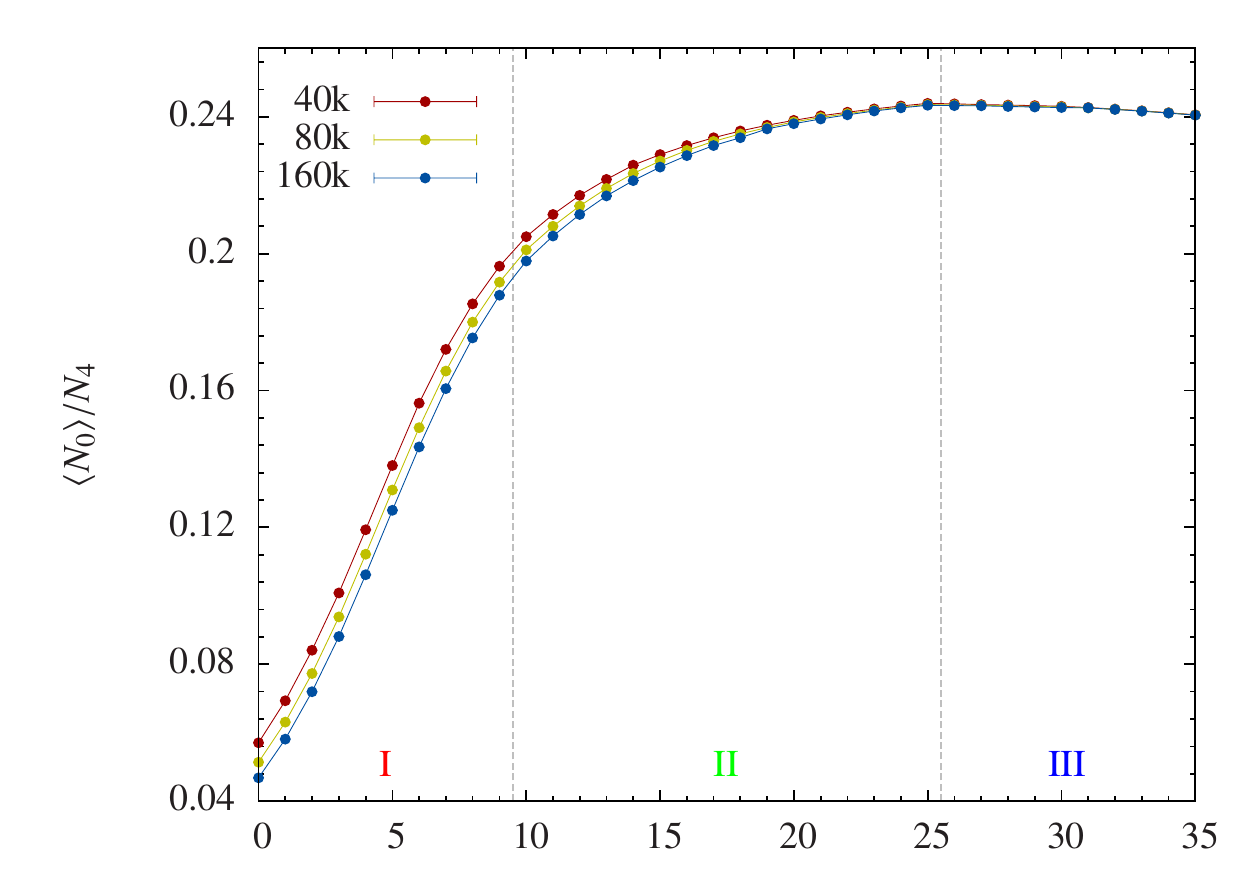}
\includegraphics[width=0.49\textwidth]{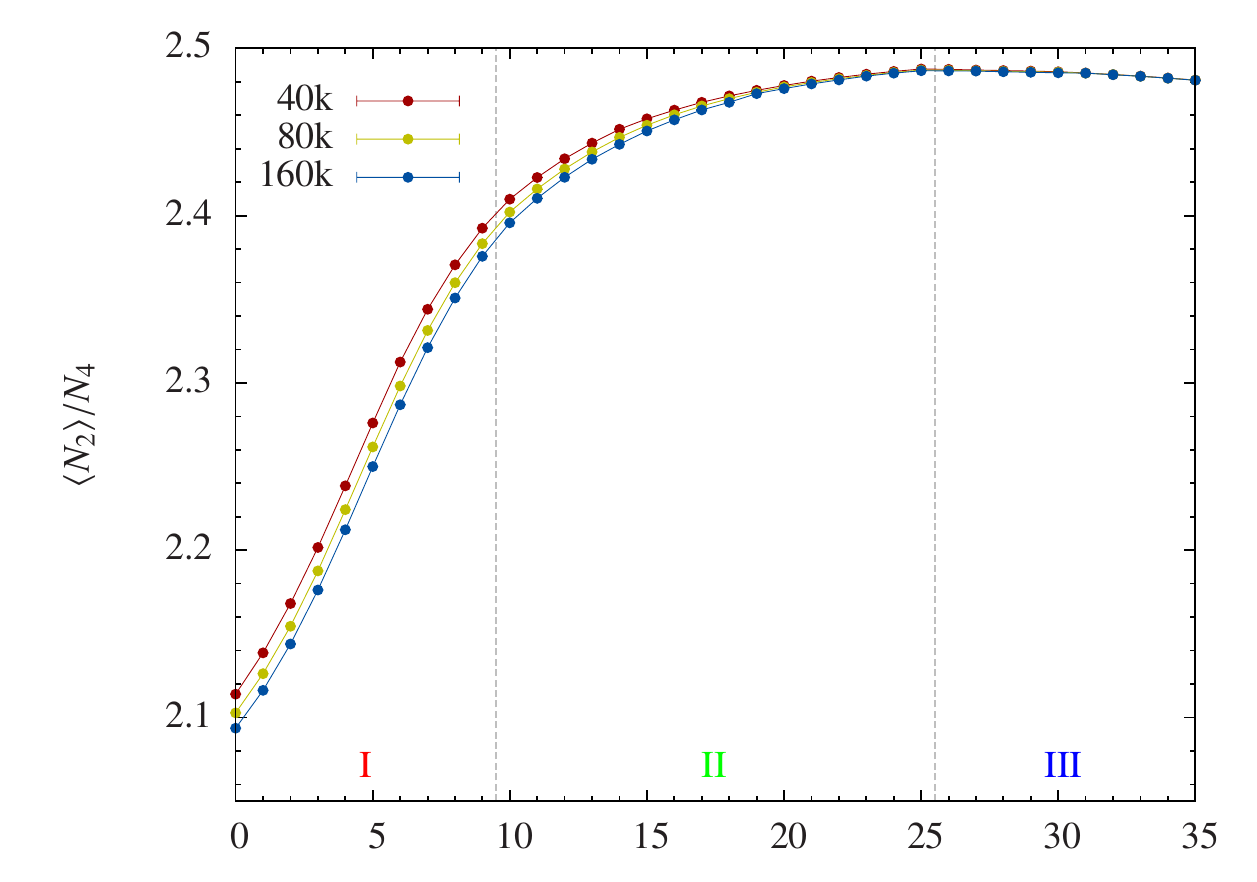}
\end{center}
\caption{Plot of $\langle N_0 \rangle / N_4$ (left) and 
$\langle N_2 \rangle / N_4$ (right) for points along the path. 
Successive points of the path are on the $x$-axis, 
the colors of the $x$-axis correspond to the colors of the path segments.}
\label{fig:pathn}
\end{figure}

The basic observables, the 
scaled average number of vertices 
$\langle N_0 \rangle / N_4$ and triangles $\langle N_2 \rangle / N_4$ 
are shown in Fig. \ref{fig:pathn}.
The successive points on the path are presented on the $x$-axis
and we have indicated the separation of the line segments I, II and III
by vertical lines. 

We do not observe any jump of 
$\langle N_0 \rangle$ or $\langle N_2 \rangle$ 
on the path between the crumpled phase and crinkled region.
There is also no jump between the branched polymer phase and 
crinkled region, in contrast to  what happens at 
$\beta = 0$ when one moves from the crumpled phase to the branched 
polymer phase. However, the scaling with $N_4$ changes 
exactly at the transition between the crinkled and the 
branched polymer phase. 
Inside the branched polymer phase 
$\langle N_0 \rangle \propto N_4$, while this scaling does not hold outside.
When we are outside the branched polymer phase  
curves corresponding to  different spacetime volumes $N_4$  
do no longer coincide,  as 
can be seen most clearly on the left side of Fig.\ \ref{fig:pathn}.

\begin{figure}[h]
\begin{center}
\includegraphics[width=0.49\textwidth]{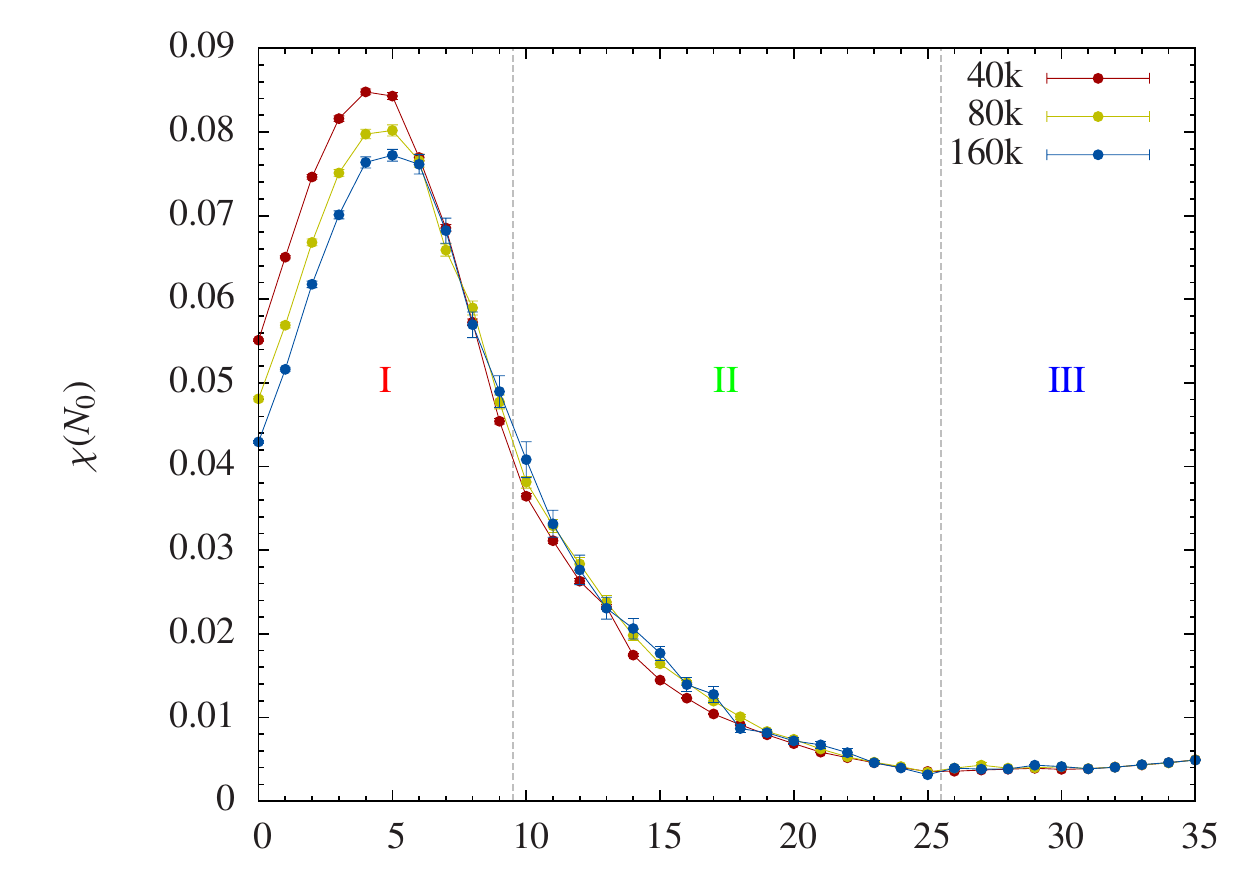}
\includegraphics[width=0.49\textwidth]{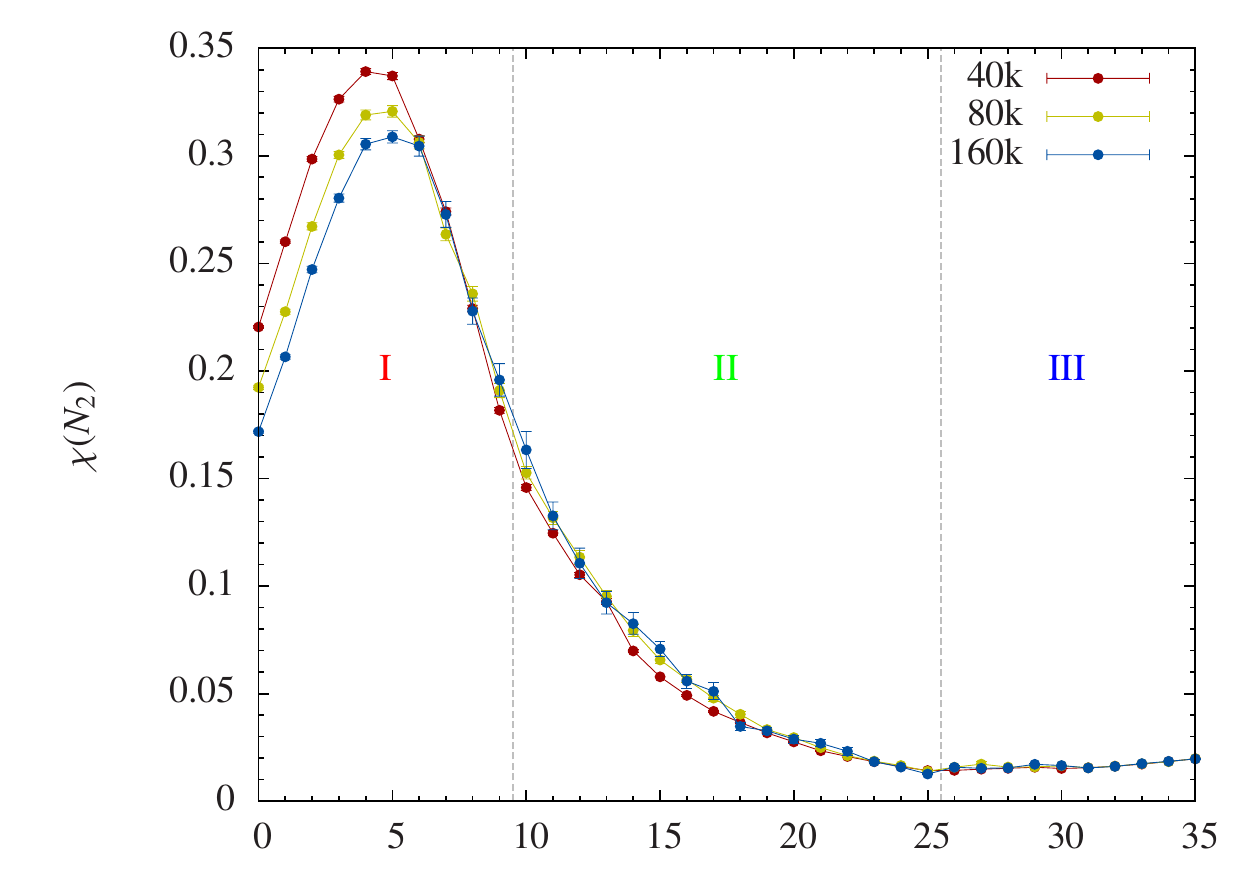}
\end{center}
\caption{Plot of variances 
$\chi( N_0 ) \equiv ( \langle N_0^2 \rangle - \langle N_0 \rangle^2) / N_4 $ 
(left) and $\chi( N_2 )$ (right) for points along the path.}
\label{fig:pathchi}
\end{figure}

Fig. \ref{fig:pathchi} shows the  measured 
susceptibilities $\chi( N_0 )$ and $\chi( N_2 )$.
Following the path, there is a peak in the  susceptibility,
located in the red segment.
It can  also be seen on the grid plot (left plot of Fig. \ref{fig:grid}).
However, the peak is decreasing with the total volume $N_4$ and can 
thus not be viewed as signaling a  first or second order 
transition between the crumpled phase and a hypothetical crinkled phase.

In addition there is a small peak 
- not well visible in Fig. \ref{fig:pathchi} - 
at the border between branched polymer phase and the crinkled 
region, being a 
remnant of a pronounced peak at $\beta = 0$. By itself it 
would be difficult to claim  that this little peak 
signals a phase transition between the branched polymer phase and 
the crinkled region. However, as we will show below, there 
are other observables which behave discontinuously precisely 
at that point.


\subsection{Triangle order $\boldsymbol{o_t}$}

\begin{figure}[h]
\begin{center}
\includegraphics[width=0.49\textwidth]{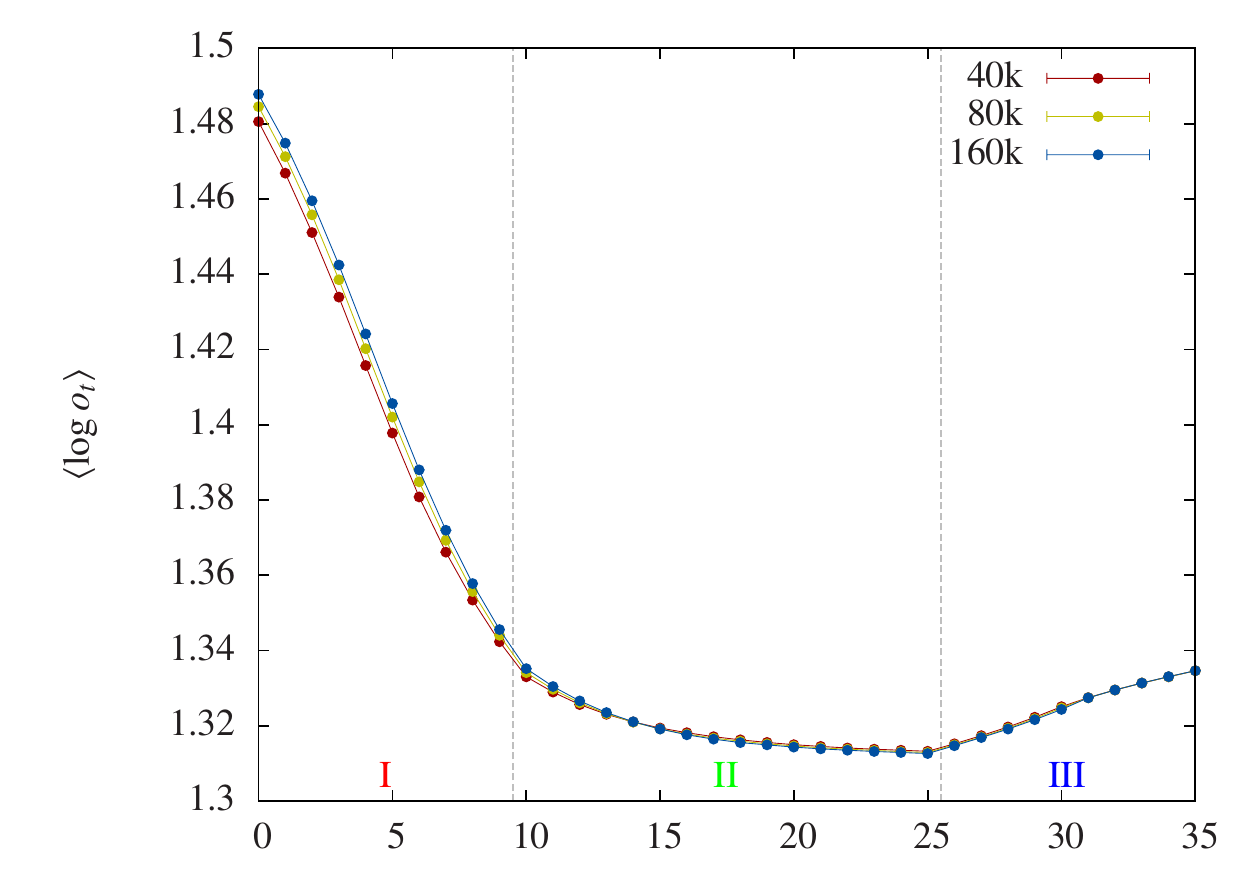}
\includegraphics[width=0.49\textwidth]{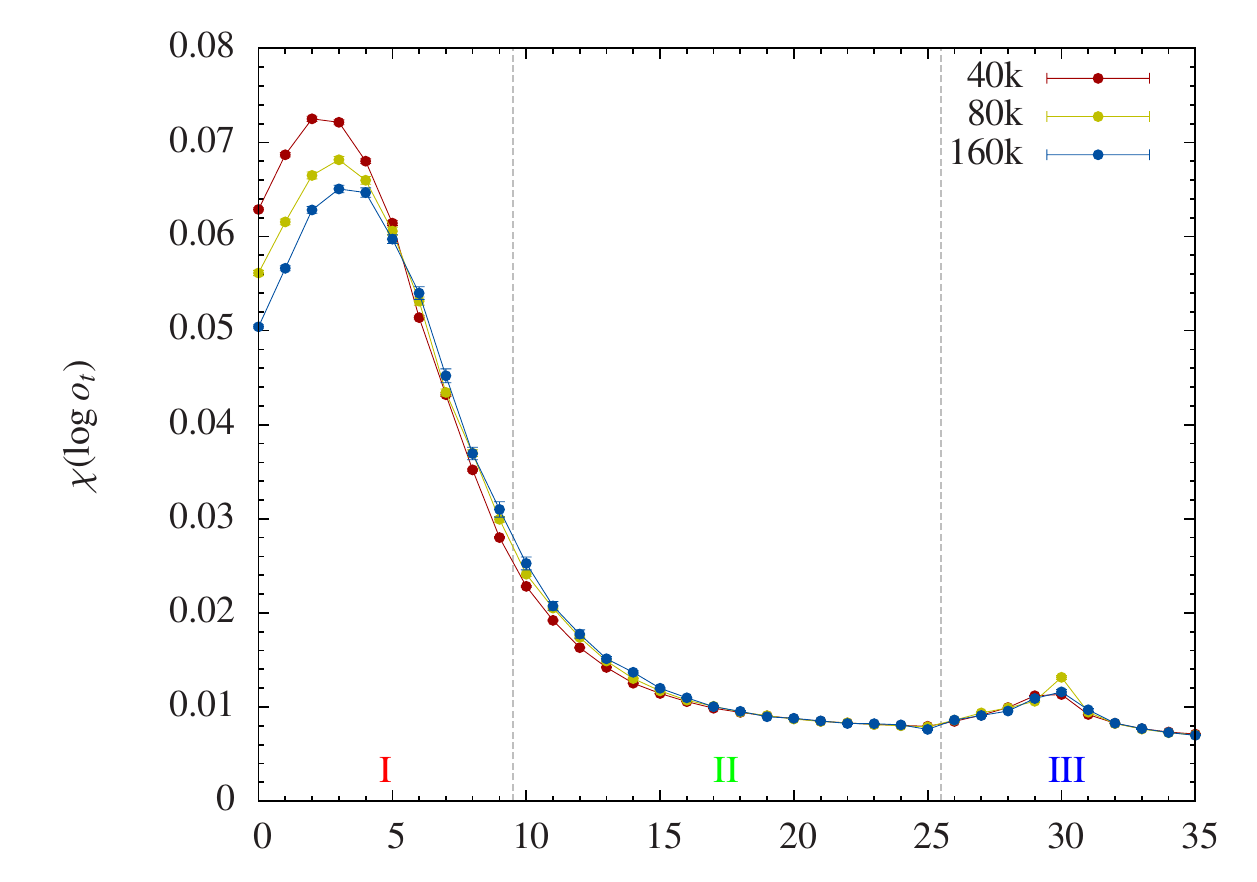}
\end{center}
\caption{Left figure: plot of the average $\langle \log o_t \rangle$
for points along the path ($o_t$ is the order of  triangle $t$). Right figure:
plot of the variance   $\chi(\log o_t) \equiv
\la (\log o_t)^2 \ra - \langle \log o_t \rangle^2$ for points along the path.}
\label{fig:pathlogot}
\end{figure}

Fig. \ref{fig:pathlogot} presents a plot of average (left) and 
variance (right) of $\log o_t$ for different total volumes $N_4$.
Because $\langle \log o_t \rangle$ is conjugate to $\beta$, 
it increases when $\beta$ increases (red and blue segments).
As for  $\chi( N_0 )$ also 
$\chi(\log o_t)= \la(\log o_t)^2 \ra - \langle \log o_t \rangle^2 $ 
has its maximum in the red 
segment, but again as for  $\chi( N_0 )$ it decreases with total volume,
and thus does not signal a second or first order 
transition between the crumpled phase and a possible crinkled phase.
There is finally a (small) peak of the variance at the transition to the 
branched polymer phase, again as for $\chi(N_0)$.

\subsection{$\boldsymbol{\la r \ra}$ and size of baby universes}

\begin{figure}[h]
\begin{center}
\includegraphics[width=0.49\textwidth]{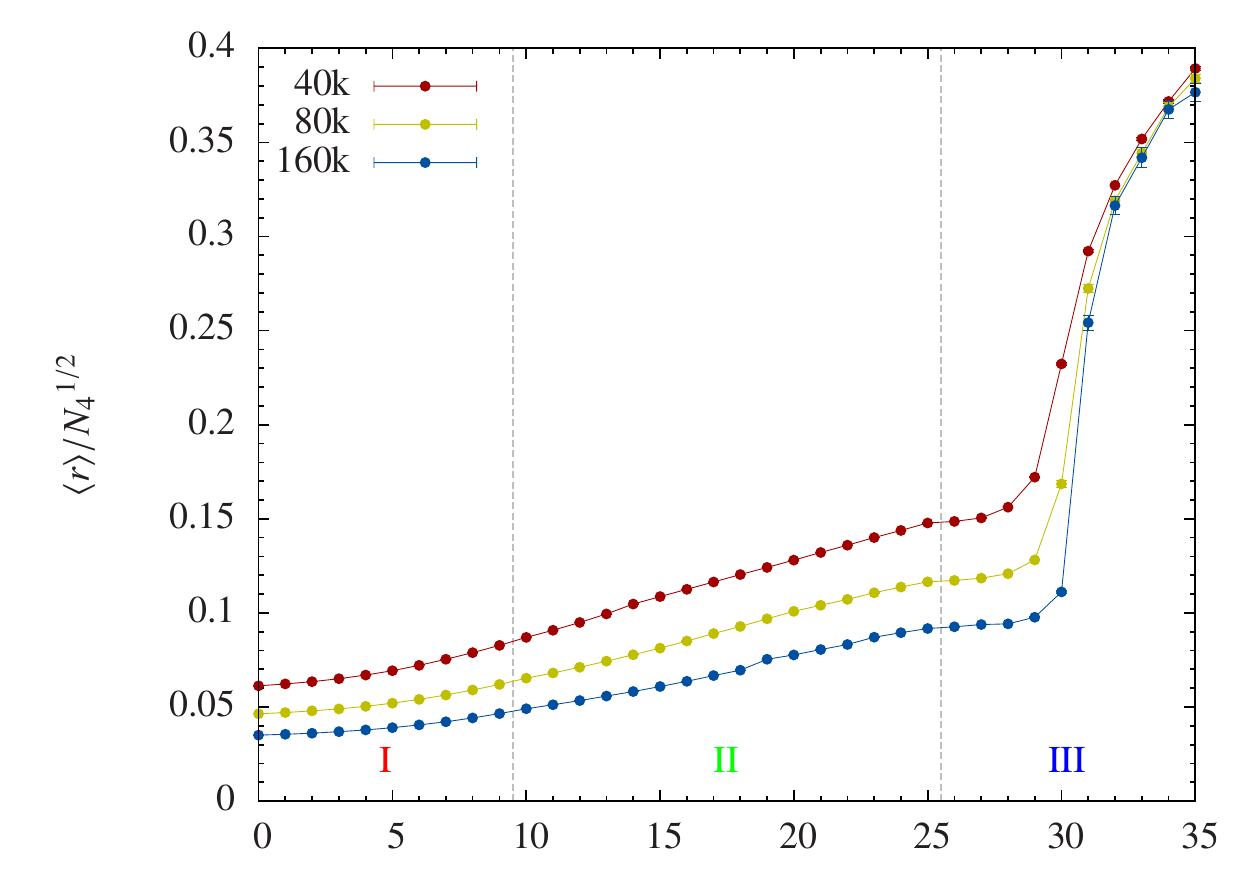}
\includegraphics[width=0.49\textwidth]{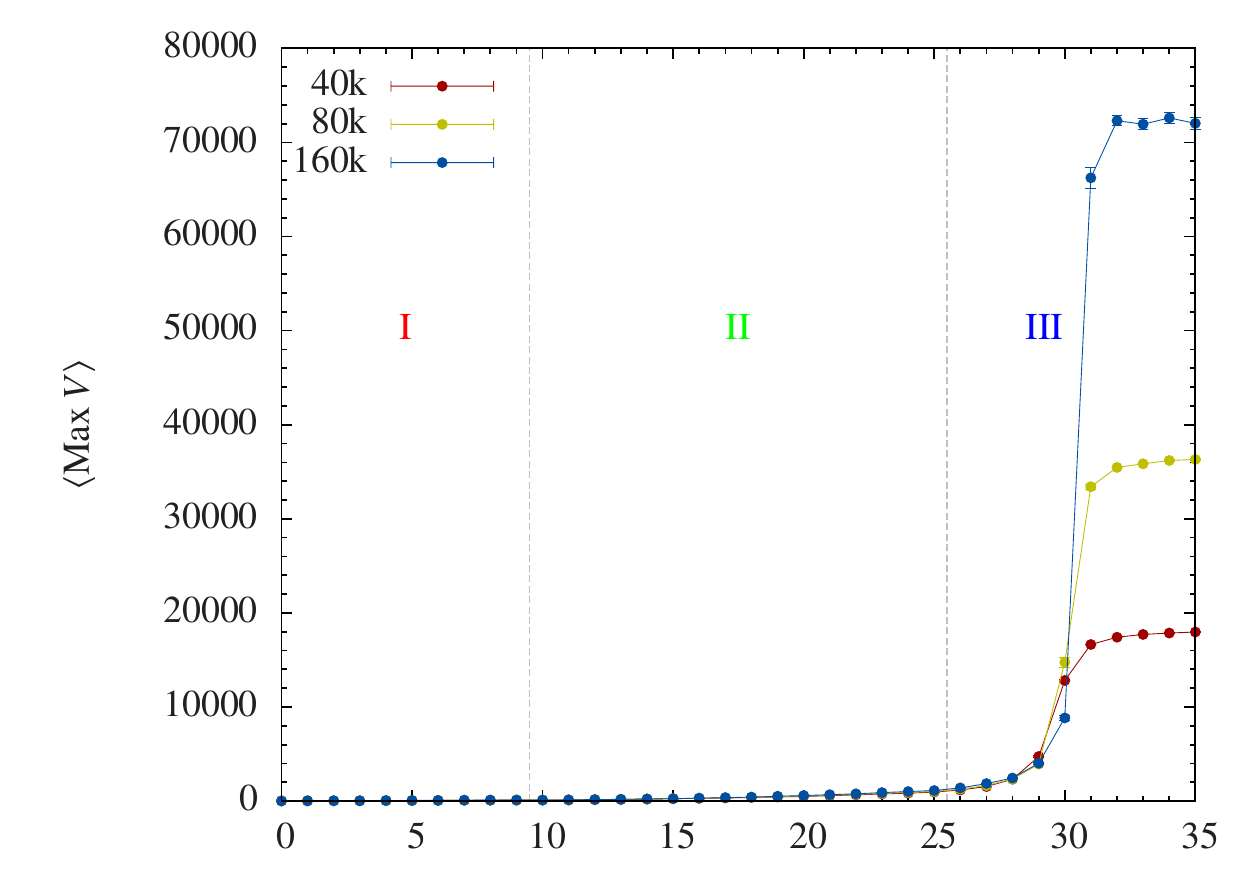}
\end{center}
\caption{Left figure: plot of $\langle r \rangle / N_4^{1/2}$
for points along the path. Right figure: 
average size of the largest baby universe for points along the path.}
\label{fig:pathavr}
\end{figure}

In the branched polymer phase, 
the Hausdorff dimension $d_h = 2$ and the average radius scales as 
$\langle r \rangle \propto N_4^{1/2}$ \cite{aj1}.
As shown in Fig.\ \ref{fig:grid} and Fig.\ \ref{fig:pathavr}, 
in this phase $\langle r \rangle$ is relatively large.
The jump of $\langle r \rangle$ at the boundary of the 
branched polymer phase is a clear signal of a phase transition.
Fig. \ref{fig:pathavr} shows that the jump of 
$\langle r \rangle$ becomes sharper as the total volume $N_4$ increases.
There is no sign of any transition between the crumpled phase 
and a possible crinkled phase.


The structure of baby universes allows us to 
extract further information about the geometry of a typical 
configuration.
Following the path from the crumpled phase to the crinkled region,
we observe the baby universe graphs dissolve gradually, starting out  
as one huge ``parent-universe'' decorated with minimally small 
baby universes  
(left graph of Fig. \ref{fig:tree}), then developing into a connected
structure without a distinct parent-universe, but with many loops
(middle graph of Fig. \ref{fig:tree}),
these loops being associated with triangles of high order.
Although the baby universe structures are very 
different in the crumpled and crinkled regions,
we do not observe any abrupt change.
When approaching the branched polymer phase, 
the loops - and high order triangles - disappear,
and a tree-like fractal structure emerges (right graph of Fig. \ref{fig:tree}).

Each minimal neck splits a triangulation into two parts.
The smaller part is what we have denoted a baby universe.
In the branched polymer phase almost surely a  minimal neck exists
which splits a configuration into two parts of nearly equal size.
Thus, the average size of the largest baby universe is very large 
and close to half of the total volume.
However, the situation is very different for typical configurations 
in the crumpled  and in the crinkled regions.
Fig. \ref{fig:pathavr} (right) shows the average size of the 
largest baby universe for successive points of the path. This is 
maybe the clearest signal of a first order transition.

\vspace{24pt}

\subsection{The Hausdorff dimension}

The Hausdorff  dimension reflects certain fractal 
structures of spacetime.  It has been studied intensively 
in two-dimensional quantum gravity  where one 
can compare numerical and analytical results, and it has been 
measured in the numerical studies of higher dimensional 
quantum gravity already referred to above.
It has a natural definition  on geometries defined by discrete triangulations
and in this sense it is an ideal observable to use in the present setup.

Let us start with an arbitrary four-simplex in our triangulation.
The neighboring four-simplices are said to have distance one to 
our chosen four-simplex. Continuing this way we can define
the spherical shell at distance $r$ from our four-simplex (note that 
the so defined spherical shell does not need to be connected). The
radial volume, i.e.\ the number of four-simplices 
in the spherical shell at distance $r$, is denoted $V(r)$, as mentioned 
earlier. We define $d_h$, the Hausdorff dimension, as the (assumed) 
power like behavior of $V(r)$:
\beq\label{jh1}
V(r) \propto r^{d_h-1},~~~1\ll r\ll N_4^{1/d_h}.
\eeq
We only expect this relation to be true as 
an ensemble average, i.e.\ if we average 
many different geometries with the appropriate 
weight coming from the action. Further, we usually 
average over the starting four-simplices. For a fixed $N_4$
we have corrections to \rf{jh1} and it is often assumed 
that one can write
\beq\label{jh2}
\la V(r) \ra_{N_4} \sim N^{1-1/d_h} v(x),~~~x=  \frac{r}{N_4^{1 / d_h}},
\eeq
where 
\beq\label{jh3}
v(x) = x^{d_h-1} F(x),~~~~F(0) =1.
\eeq
Formulas \rf{jh2} and \rf{jh3} have the form of finite size scaling 
relations and are convenient to use when trying to determine $d_h$.
Note that a consequence of the assumed scaling is that 
\beq\label{jh4}
\la r \ra_{N_4} \propto N_4^{1/d_h}.
\eeq
 
\begin{figure}
\begin{center}
\includegraphics[width=0.49\textwidth]{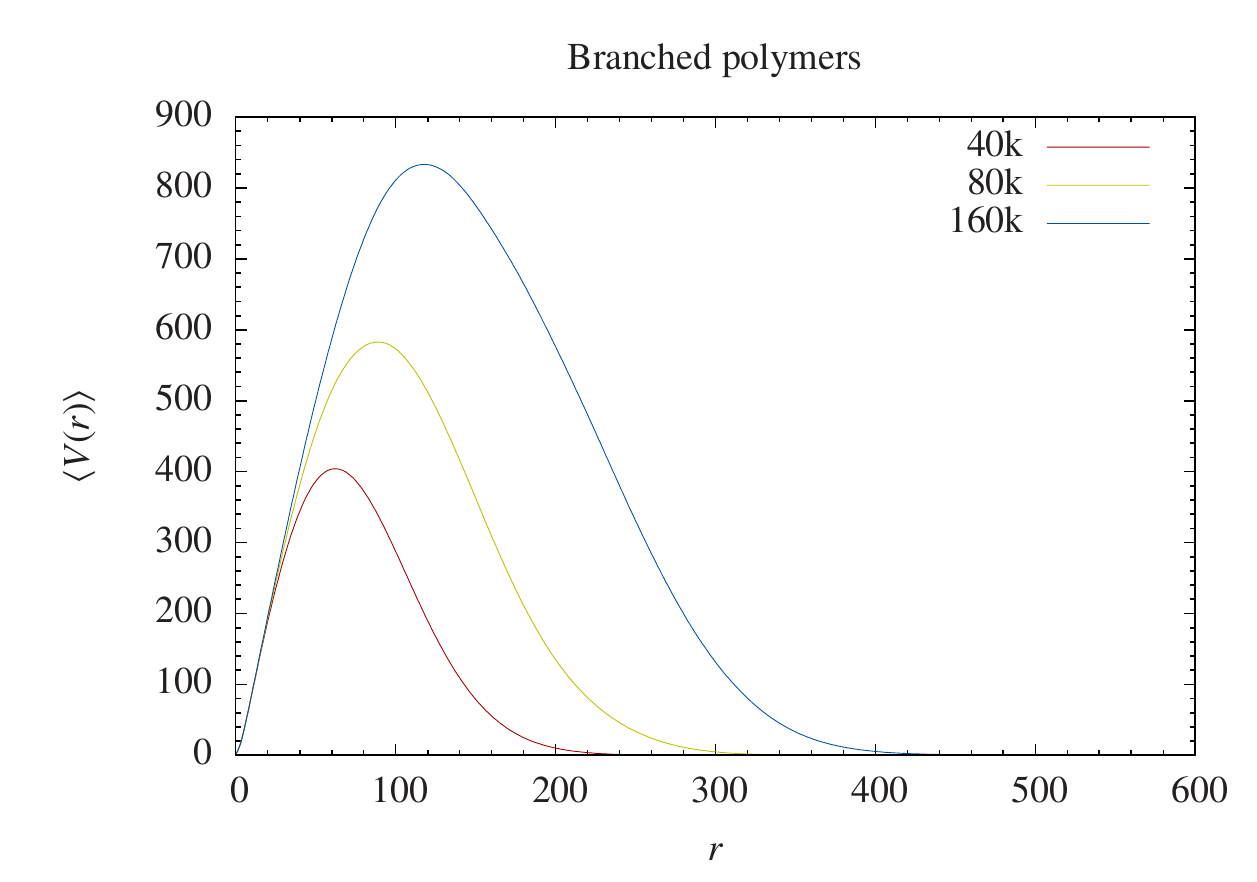}
\includegraphics[width=0.49\textwidth]{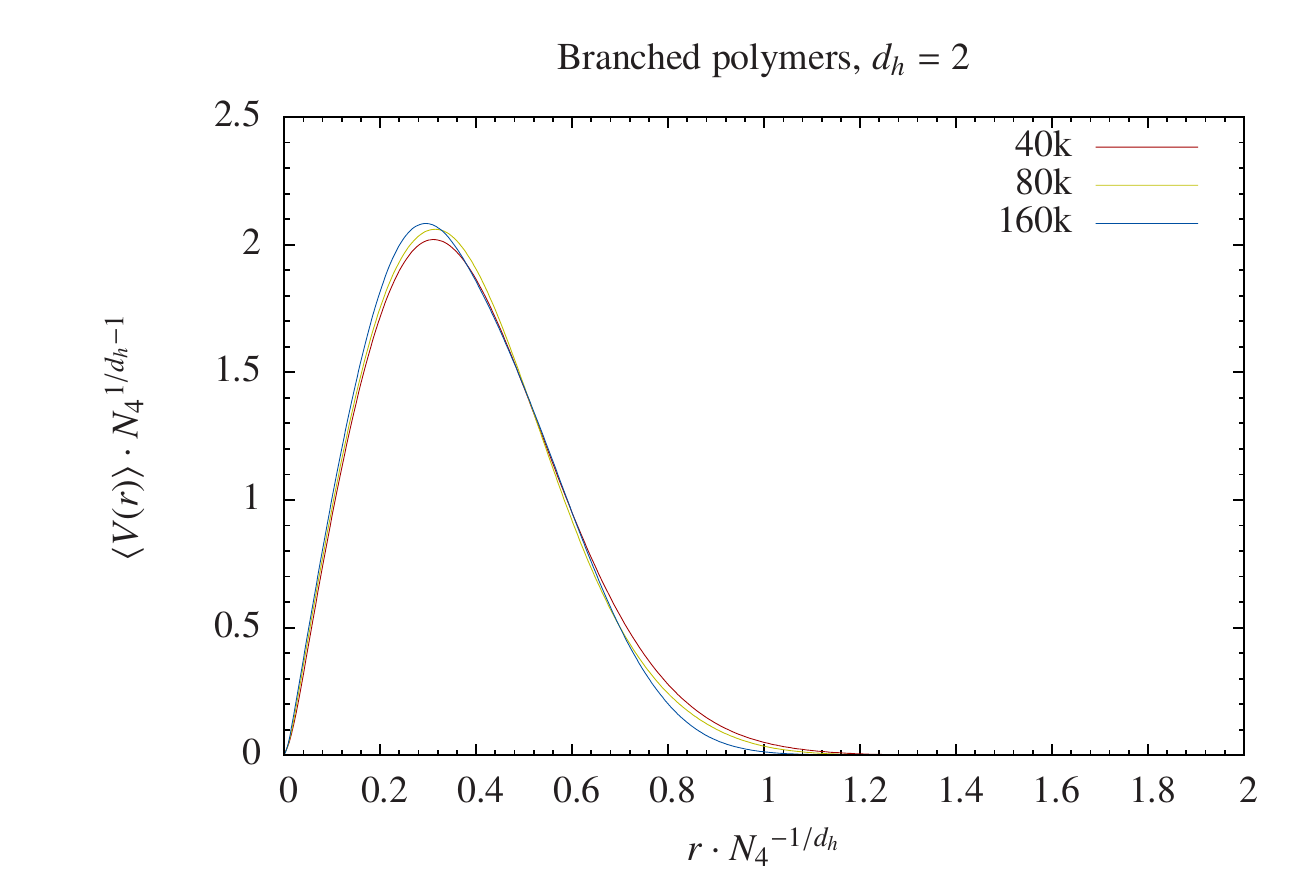}
\end{center}
\caption{$V(r)$ and $v(x)$ in the branched polymer phase 
for $N_4= 40\rk$, $80\rk$ and $160\rk$.}
\label{BP}
\end{figure}

Let us describe the results of the measurements of the Hausdorff 
dimension $d_h$. Everywhere in the branched polymer phase we find nice 
agreement with  scaling assumptions \rf{jh2} and \rf{jh3},
and the data are consistent with $d_h=2$, the result for 
branched polymers. This is in agreement with old results obtained 
along the line $\beta =0$ in the branched polymer phase.
In Fig.\ \ref{BP} we have shown the result of such a finite size
scaling for the choice $d_h=2$. One can refine the analysis and 
determine $d_h$ with reasonable accuracy to be two, but since 
this is not too important for the discussion we skip the details. 

In the crumpled and crinkled regions of the phase diagram the 
scaling \rf{jh2} and \rf{jh3} are not well satisfied 
and cannot be used to determine a $d_h$ with any precision.
This is in agreement with the old observations along the 
$\beta=0$ part of the crumpled region, where it was judged
that the Hausdorff dimension was very large since 
the configurations were centered around two neighboring vertices of 
order $N_4$ and the linear extension did hardly change with $N_4$. 
 Let us follow
the path on Fig.\ \ref{fig:phasediagram} from the crumpled phase, 
starting at $\beta=0$ and moving towards
the crinkled region. As already emphasized there is 
no observed phase transition between the crumpled region and the crinkled 
region. This is also the case when it comes to the Hausdorf dimension.
As mentioned, it starts out large at $\beta = 0$. Moving into 
the crinkled phase the structure of the two singular neighboring 
vertices is resolved and the extensions
of typical configurations grow. Although \rf{jh2} and \rf{jh3} 
are  not well satisfied we have found another way to estimate
$d_h$.    Surprisingly, the average radial profile is almost  
symmetric with respect to the reflection $V(r) \rightarrow V(R - r)$.
Thus, before performing the average over configurations 
one can \emph{center} the volume profiles
using following procedure. We find \emph{the center of mass} 
or the average radius of the volume profile $V(r)$, defined as 
\beq\label{jh5}
 r_{av}  = \frac{1}{N_4} \sum_r r \cdot V(r), 
\eeq
and redefine the radius coordinate $r \rightarrow r -  r_{av}$ so that
the \emph{center of mass} is located at $r = 0$.
Afterwards, we perform the average over configurations 
and find the  value of $d_h$ for which the scaled 
profile $v_{cm}(x)$ becomes volume independent.

\begin{figure}
\begin{center}
\includegraphics[width=0.49\textwidth]{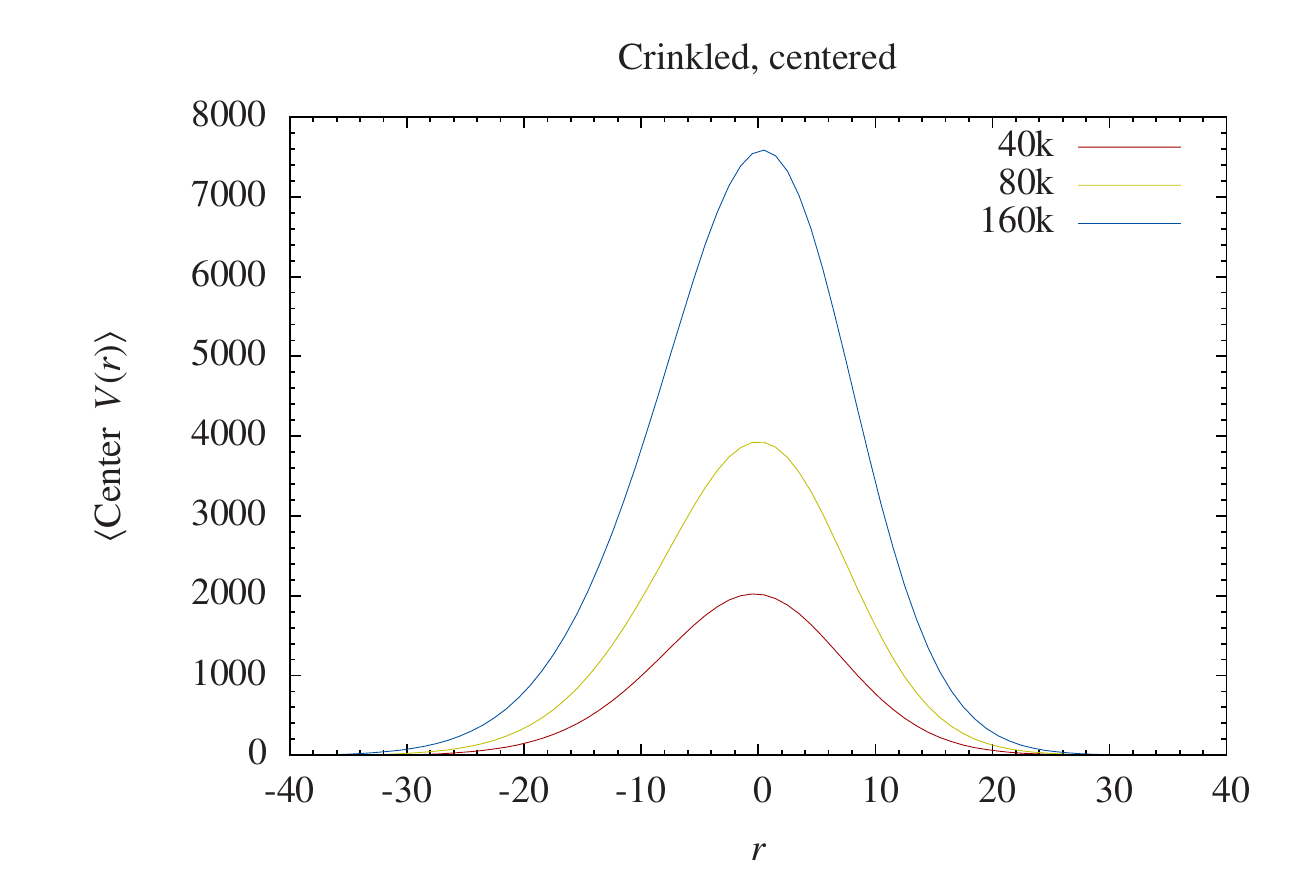}
\includegraphics[width=0.49\textwidth]{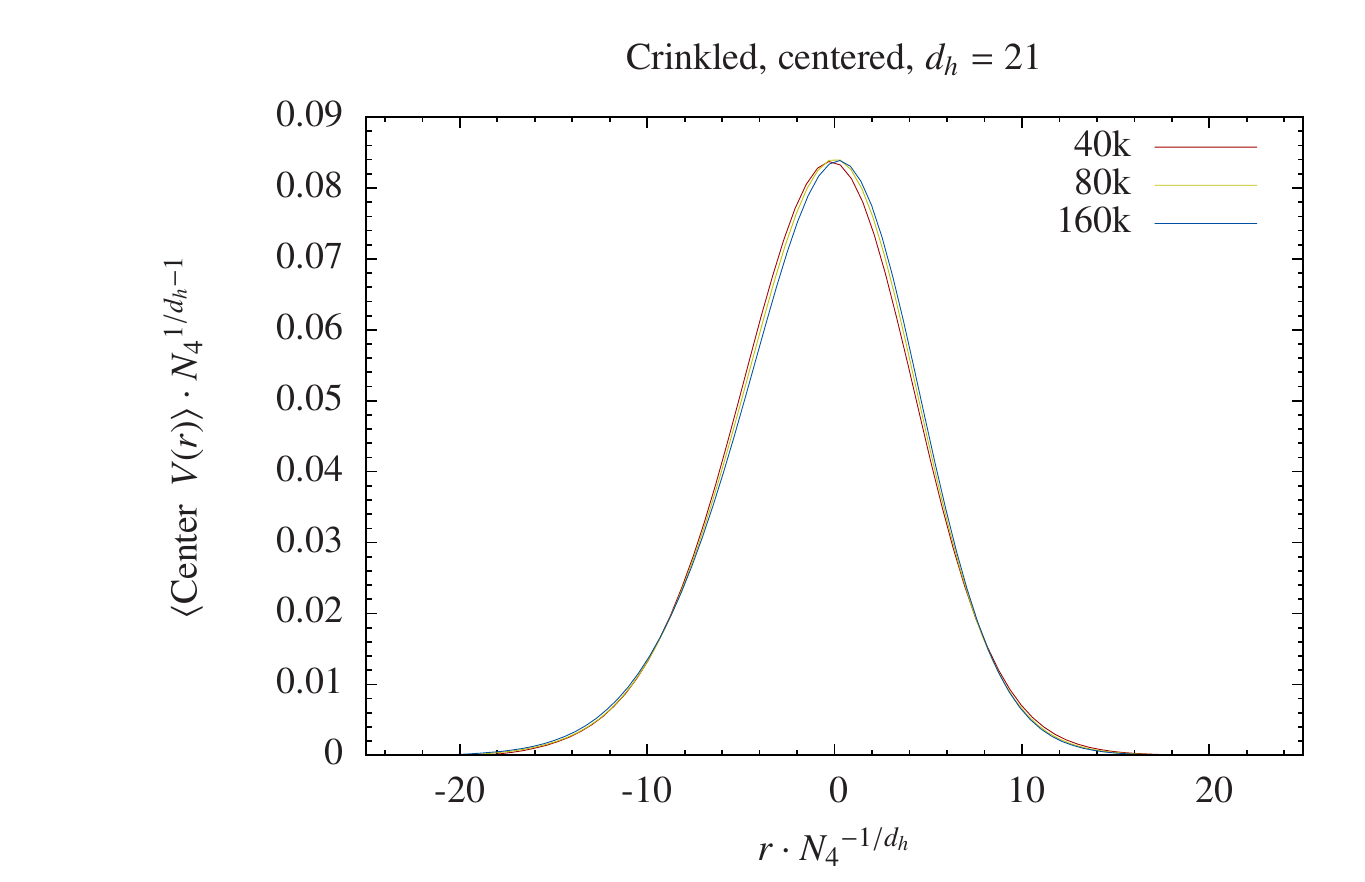}
\end{center}
\caption{$V_{cm}(r)$ and $v_{cm}(x)$ in the crinkled region 
for $N_4= 40\rk$, $80\rk$ and $160\rk$.}
\label{fig:hausdorff}
\end{figure}

The centered radius profiles $V_{cm}(r)$ and the corresponding scaled and 
centered  radius volume profiles $v_{cm}(x)$ are shown in 
Fig.\ \ref{fig:hausdorff}  for $N_4 = 40\rk, 80\rk, 160\rk$ 
for a choice of coupling constants in the  crinkled region.
Although the configurations in the crinkled region are not so strongly 
collapsed as in the crumpled region $d_h$ still comes out very high 
( $d_h \approx 21$).
Such large values of $d_h$ may indicate that in the infinite volume limit
the Hausdorff dimension is infinite. To estimate $d_h$ more precisely 
one would clearly need larger values of $N_4$. However, the result
clearly differs from the $d_h$ in the branched polymer phase and
is much closer to the results obtained in the crumpled region.

\subsection{The spectral dimension}

\begin{figure}
\begin{center}
\includegraphics[width=0.49\textwidth]{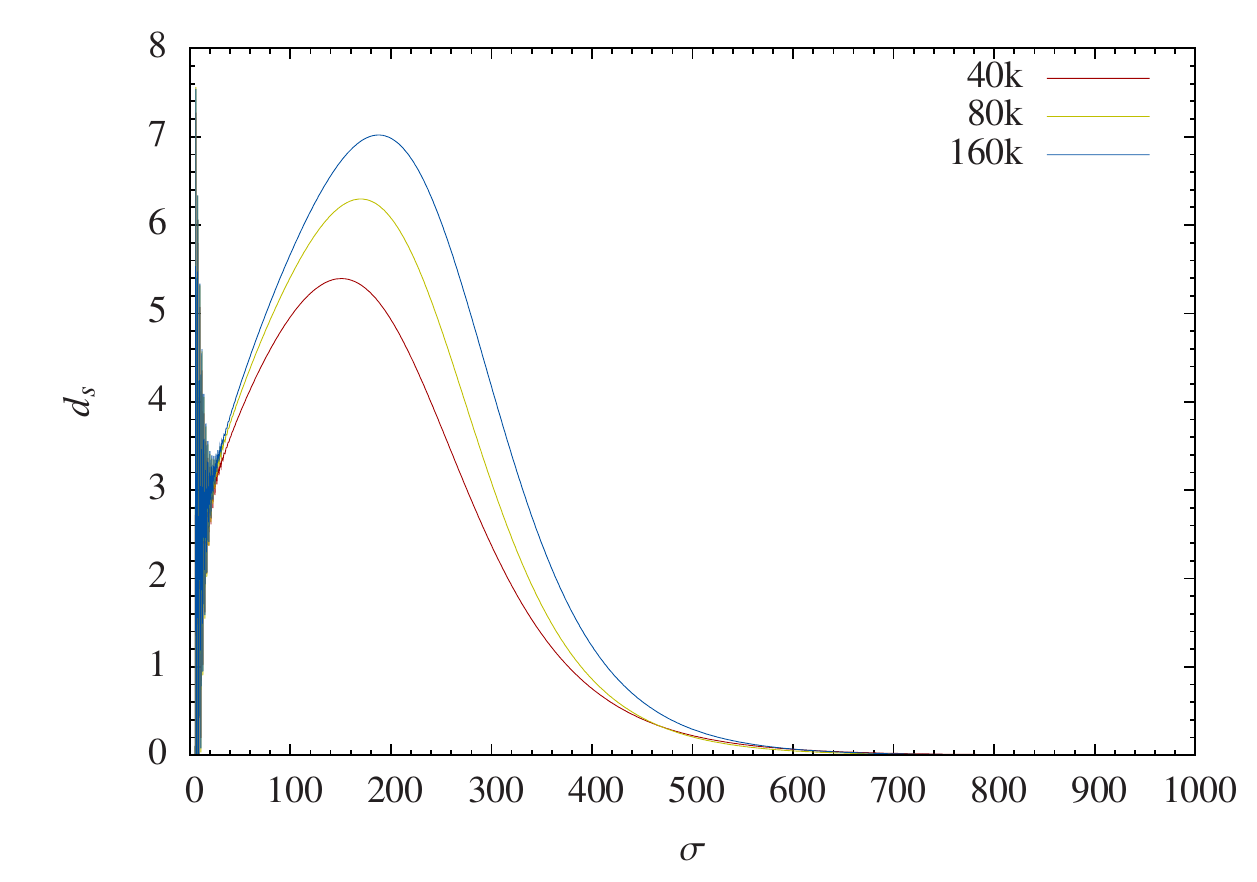}
\includegraphics[width=0.49\textwidth]{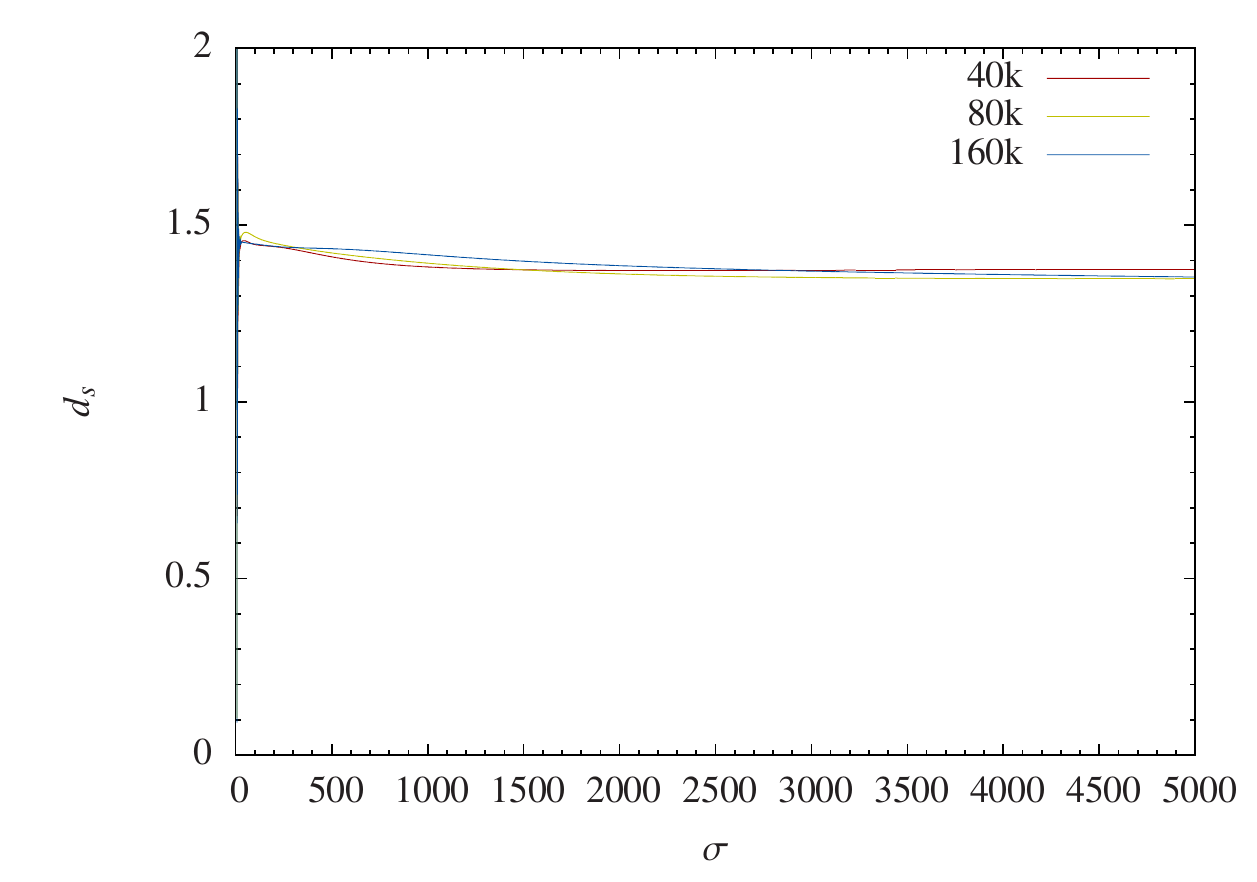}\\
\includegraphics[width=0.80\textwidth]{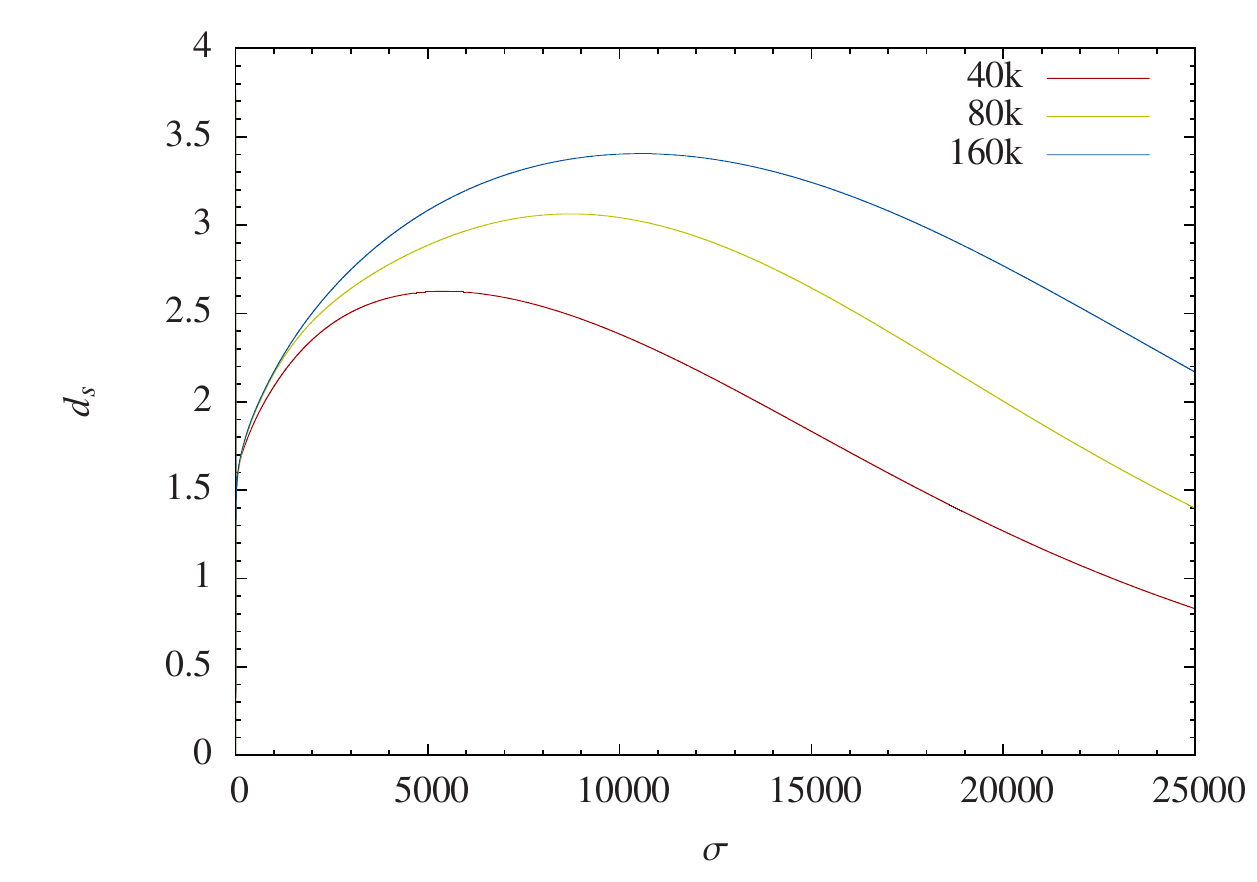}\\
\end{center}
\caption{
Spectral dimension $d_s$ as a function of diffusion time $\sigma$.
Left: crumpled phase ($\kappa_2 = 0.5, \beta =0.0$). 
Right: branched polymers phase($\kappa_2=2.0, \beta =-1.0$). 
Bottom: crinkled region ($\kappa_2=2.0, \beta =-2.0$).}
\label{fig:spectral}
\end{figure}

The work reported in this article was triggered by the 
interesting measurements of the spectral 
dimension reported in \cite{Laiho}. Let us turn to the measurement 
of the spectral dimension for our ensemble of 
quantum geometries. It can be extracted by studying a 
diffusion process on the given ensemble of geometries. It shares 
with the Hausdorff dimension the nice property that it can be defined 
on piecewise linear geometries in a simple way. We will study the 
diffusion of a particle, performing a random walk between (the centers of)
neighboring four-simplices. Denote by   $\rho(i, i_0; \sigma)$ the 
probability that a particle starting at simplex $i_0$ is 
found at simplex $i$ after the fictitious (discrete) diffusion time $\sigma$.
$\rho(i, i_0; \sigma)$ satisfies the following discrete diffusion equation:
\beq\label{jd1} 
\rho(i, i_0; \sigma + 1) = 
\frac{1}{5} \sum_{j \leftrightarrow i} \rho(j, i_0; \sigma), 
\quad \quad \rho(i, i_0; 0) = \delta_{i\,i_0}, 
\eeq
where the sum is evaluated over all simplices $j$ adjacent to $i$.
Eq.\ \rf{jd1} expresses that the particle performs a random walk,
jumping between centers of neighboring four-simplices.  
The \emph{average return probability},
\beq\label{jd2} 
P(\sigma) = 
\left\langle \left\langle \rho(i_0, i_0; \sigma) \right\rangle_{i_0} 
\right\rangle_{conf}, 
\eeq
describes the probability of finding a particle 
at the initial point after diffusion time $\sigma$.
The inner average is performed over initial simplices $i_0$.
The outer average is performed over configurations.

Let us define the spectral dimension $d_s(\sigma)$ as
\beq\label{jd3} 
d_s(\sigma) \equiv -2 \frac{\dd \log P(\sigma)}{\dd \log \sigma}.
\eeq
For diffusion on $R^d$ the spectral dimension is equal to $d$ and 
independent of (the continuous) diffusion time $\sigma$. 
If we consider a smooth compact manifold $d_s$ will be a function 
of $\sigma$ which in the limit where $\sigma \to 0$ is equal to 
the topological dimension of the manifold and which in the limit 
where $\sigma \to \infty$ goes to zero.
For diffusion on piecewise linear manifolds as defined here, the 
short time diffusion reflects the discretization used. Typically one 
can obtain quite different results for even and odd discretized  
times if one uses the simple implementation \rf{jd1} for the diffusion.
However, usually after some diffusion time has passed
one has $d_s(\sigma_{odd}) \approx d_s(\sigma_{even})$ and for $\sigma$ not 
too large there is a plateau independent of $\sigma$ which 
we can then identify with {\it the} spectral dimension $d_s$. After 
that, for a finite $N_4$, the spectral dimension will decrease slowly 
to zero. 

In Fig.\ \ref{fig:spectral} we have shown 
the spectral dimension as a function of diffusion time $\sigma$ 
in the crumpled, crinkled and branched polymer regions. 
The values of  $N_4$ used are  $40\rk, 80\rk, 160\rk$. 
For $\sigma < 50$ lattice artifacts are 
pronounced but for larger values $d_s(\sigma_{odd}) \approx d_s(\sigma_{even})$
merge into a smooth curve.

In the branched polymer phase we see the  plateau mentioned above (and 
we have not run the diffusion process long enough to see $d_s \to 0$).
The value of $d_s$ is close to 4/3, the theoretical value for branched
polymers, again providing evidence that the configurations indeed 
are very much like branched polymers, despite being four-dimensional 
triangulations.

In the crumpled phase we see no plateau at all and clearly 
the maximum is increasing with $N_4$ and we observe a rapid drop 
towards zero after the maximum. This reflects the very short 
distances available for diffusion despite 
the large values of $N_4$ and thus effectively the high dimensionality of 
the configurations. If one can talk about a 
spectral dimension at all it is clearly large.

In the crinkled region the behavior of the spectral dimension is 
somewhat similar to what we observed in the crumpled region, only the 
maxima of $d_s(\sigma)$  are somewhat smaller and the diffusion time during 
which $d_s(\sigma)$  is different from zero is longer.
This is a reflection of the larger
extention of the configurations in the crinkled regions for a 
given $N_4$. However, the important message is really that 
the maximum of $d_s(\sigma)$ shows no sign of converging 
as a function of $N_4$. This is in contract to the situation
in four-dimensional CDT, where one also observes 
a $\sigma$ dependent $d_s$, but as a function of $N_4$ the 
curves $d_s(\sigma)_{N_4}$ converge to a universal 
curve $d_s(\sigma)_{N_4=\infty}$. We cannot rule out that 
the same could happen here for very large $N_4$, but from
the present data we cannot identify anything like 
a universal $d_s(\sigma)_{N_4=\infty}$.

\section{Conclusions}

As described in the Introduction, introducing $\bt$ as 
an additional coupling constant in DT-regularized Euclidean Quantum
gravity is potentially very interesting. It could 
unite a number of different approaches to quantum gravity:
the DT lattice approach, the higher curvature approach leading 
to asymptotic freedom and the asymptotic safety approach based 
on the existence of a non-Gaussian UV fixed point. 
It could also, in principle, make connection to the CDT lattice approach
since at least the spectral dimension in the crinkled phase was reported in 
\cite{Laiho} to have a scale dependence 
similar to the one found in the CDT lattice approach to quantum gravity.
However, at least applying conventional wisdom, 
in order to be interesting from a continuum point of view 
one has to be able to localize a phase transition point where 
continuum physics is recovered and a whole number of lattice
artifacts fade away. Unfortunately we have {\it not} been 
able to observe such a phase transition point. What we {\it have} 
observed is a first order phase transition line which is a natural 
continuation of the first order phase transition between the crumpled 
and the branched polymer phase observed originally at $\bt=0$. 
Such a continuation was of course expected when we explored the 
$(\kp_2,\bt)$ coupling constant plane, but it could have  changed 
into a second order transition point {\it if} there had been 
a genuine crinkled phase and a phase transition between the 
crinkled and the crumpled phases. However, we do not observe any signal, 
growing with the total volume, of a phase transition 
between the crumpled phase and the crinkled phase.
Configurations in the crinkled region look less ``crumpled'' 
($V(r)$, minbu trees, spectral dimension),
but the change is gradual when 
receding from the crumpled phase and it seems to be a finite size effect.

While the results reported here are negative results, we nevertheless
feel that they are important in the sense that 
they shown that one should probably not spend more
time investigating the so-called crinkled phase. 
As discussed in the Introduction, 
there should exist an asymptotically free--asymptotically safe  
Euclidean ``gravity'' theory,
obtained by adding higher curvature terms which serve to make the 
theory renormalizable and at the same time cure the unboundedness 
problem of the Euclidean Einstein-Hilbert action. This might 
not be the gravity theory we want, and if it could in some way be 
rotated back to spacetime with Lorentzian signature it might 
not be unitary, but it should exist. Thus we should be able 
to identify it in the DT lattice approach, {\it provided} 
we can find a decent way to implement the higher curvature 
terms in the DT formalism. The present results indicate that the 
attempts to use the Regge curvature \rf{1.1}, even in some 
more general way via the suggested measure term \rf{1.6} , 
are too naive, and they 
tell us to go back to the drawing board.

\vspace{.5cm}      
\noindent {\bf Acknowledgments.} 
The authors acknowledge support from the ERC-Advance grant 291092,
``Exploring the Quantum Universe'' (EQU). JA acknowledges support 
of FNU, the Free Danish Research Council, from the grant 
``quantum gravity and the role of black holes''. JJ acknowledges the 
support of the grant DEC-2012/06/A/ST2/00389 from 
the National Science Centre Poland.  Finally this research 
was supported in part by the Perimeter Institute of Theoretical Physics.
Research at Perimeter Institute is supported by the Government of Canada
through Industry Canada and by the Province of Ontario through the 
Ministry of Economic Development \& Innovation.

\end{document}